\documentclass[
pra,
aps,
twocolumn,
superscriptaddress,
longbibliography,
floatfix,
notitlepage]{revtex4-2}

\usepackage[utf8x]{inputenc}
\usepackage[english]{babel}
\usepackage[T1]{fontenc}
\usepackage{lmodern}

\usepackage{amsmath,amsfonts,amssymb,amsthm,bm,times,dcolumn,dsfont}
\usepackage{mathrsfs}
\usepackage{microtype}
\usepackage{braket}
\usepackage{gensymb} 
\usepackage{xcolor}
\usepackage[colorlinks={true}, citecolor={blue}, filecolor={blue}, linkcolor={blue}, urlcolor={blue}]{hyperref}
\usepackage{graphicx,color}
\usepackage[caption=false]{subfig}
\usepackage{float}

\begin{document}

\title{Quantum-enhanced passive remote sensing}

\author{Emre Köse}
  \email{saban-emre.koese@uni-tuebingen.de }
\affiliation{Institut f\"{u}r Theoretische Physik, Eberhard Karls Universit\"{a}t T\"{u}bingen, 72076 T\"{u}bingen, Germany}

\author{Gerardo Adesso}
  \email{gerardo.adesso@nottingham.ac.uk}
\affiliation{School of Mathematical Sciences, University of Nottingham, University Park, Nottingham NG7 2RD, United Kingdom}

\author{Daniel Braun}
  \email{daniel.braun@uni-tuebingen.de}
\affiliation{Institut f\"{u}r Theoretische Physik, Eberhard Karls Universit\"{a}t T\"{u}bingen, 72076 T\"{u}bingen, Germany}

    \date{\today}
\begin{abstract}
We investigate theoretically the ultimate resolution that can be
achieved with passive remote sensing in the microwave regime
used e.g.~on board of satellites observing Earth, such as the Soil
Moisture and Ocean Salinity (SMOS) mission. We give a fully quantum
mechanical analysis of the problem, starting from thermal
distributions of microscopic currents on the surface to be imaged that
lead to a mixture of coherent states of the electromagnetic field
which are then measured with an array of receivers.  We derive the
optimal detection modes and measurement schemes that allow one to saturate
the quantum Cram\'er-Rao bound for the chosen parameters that
determine the distribution of the microscopic currents.  For parameters
comparable to those of SMOS, a quantum enhancement of the spatial
resolution by more than a factor of 20 should be possible with a single measurement and a single detector, and a resolution down to the order of 1 meter and less than a 1/10 Kelvin for the theoretically possible maximum number of measurements.  
\end{abstract}
\maketitle
\section{Introduction}
Optical imaging has evolved dramatically since the discovery that
Ab\'e's and Rayleigh's resolution limit comparable to the wavelength of the used light is not a fundamental bound. This was demonstrated
experimentally with a series of works starting with stimulated emission depletion in 1994 by Hell
\cite{hell_breaking_1994}, who showed that
decorating molecules with fluorophores and quenching these selectively, imaging of a molecule with nanometer resolution could be achieved in the optical
domain (see \cite{hell_farfield_2007} for a review). This was followed in 2016 by theoretical work by Tsang and
coworkers \cite{tsang_quantum_2016} who framed the problem of the ultimate resolution of two-point
sources in terms of quantum parameter estimation, a very natural approach given that quantum parameter estimation theory was originally
motivated by generalizing the classical Cram\'er-Rao bound that had
long been used in radar detection to the optical domain
\cite{helstrom_detection_1967,helstrom_quantum_1969,helstrom_cramerrao_1973,helstrom_estimation_1970}. Tsang and coworkers showed that even in
the limit of vanishing spatial separation between the two sources a
finite quantum Fisher information (QFI) for that parameter remains, whereas the classical Fisher information degrades in agreement with Rayleigh's bound \cite{tsang_quantum_2015}.  A large
body of theoretical work followed that incorporated important concepts such
as the point spread function for analyzing optical lens systems,  and
mode-engineering such as SPADE for optimal detection modes \cite{tsang_quantum_2019,zhou_modern_2019,sorelli_momentbased_2021,rehacek_multiparameter_2017,napoli_superresolution_2019,nair_farfield_2016,lupo_ultimate_2016,larson_resurgence_2018,kurdzialek_superresolution_2021,kolobov_quantum_2000,ang_quantum_2017,bisketzi_quantum_2019,bojer_quantitative_2021,datta_subrayleigh_2020,dealmeida_discrimination_2021,liang_coherence_2021,tsang_subdiffraction_2017,tsang_quantum_2015}, reminiscent of the engineering of
a ``detector mode'' for single-parameter estimation of light sources \cite{pinel_ultimate_2012}.
Experimental work in recent years validated this new approach to imaging \cite{backlund_fundamental_2018,mazelanik_optical_2021,paur_achieving_2016,pushkina_superresolution_2021}.  Optical
interferometers were investigated in
\cite{lupo_quantum_2020,bojer_quantitative_2021,gottesman_longerbaseline_2012,khabiboulline_optical_2019}. {The resolution for general parameter estimation for weak thermal sources was studied in \cite{tsang_quantum_2011}. Recently, the spatial resolution of two point sources for two mode interferometers was examined for the far field regime \cite{wang_superresolution_2021}. }

In this work, we investigate the ultimate limits of passive remote sensing in the microwave regime with a satellite of the surface of Earth. There, the state of
the art is the use of antenna arrays for synthesizing
interferometrically a large antenna with corresponding enhanced
resolution. For example, the SMOS (The Soil Moisture and Ocean Salinity) interferometer achieves a resolution of
about 35 km, flying at the height of about $758$ km and using a Y-shaped array of
69 antenna \cite{anterrieu_resolving_2004,corbella_visibility_2004,levine_synthetic_1999,thompson_interferometry_2017}. Each
antenna measures in a narrow frequency band 1420-1427 MHz with a central wavelength around $\lambda\sim 21$ cm and in real-time the electric fields corresponding to the thermal noise emitted by
Earth according to the local brightness temperatures on its surface.
The signals are filtered and interfered numerically, implementing thus
purely classical interference, which implies a resolution governed by
the van Cittert-Zernike theorem
\cite{vancittert_wahrscheinliche_1934,zernike_concept_1938,braun_generalization_2016}. Recently, it was shown
theoretically, that larger baselines can be synthesized by using the motion of the satellite but at the price of the radiometric
(i.e.~temperature) resolution \cite{braun_fouriercorrelation_2018}. The question naturally arises to what extend the resolution can be improved by
using methods of quantum metrology.  As in the optical domain
the answer can be found by analyzing the quantum Cram\'er-Rao bound and
then trying to find the optimal measurements that can achieve it. We
solve this problem in general, for an arbitrary antenna array defined
by the positions of individual antenna, in the sense of finding --- at
least numerically --- the optimal modes for measuring the electric
fields.  We go beyond the situation of localized point sources that
has become a favorite simplification in the field and describe the
sources as randomly fluctuating microscopic current distributions
which in turn generate the electromagnetic field noise, ultimately
measured by the satellite. This is closer to the literature on passive
remote sensing in the microwave regime and allows a direct comparison
with the van Cittert Zernike theorem.  We also make use of the
scattering matrix formalism introduced in this context in
\cite{jeffers_quantum_1993}. The thermal fluctuations of the microscopic
currents lead to Gaussian states of the microwave field
\cite{liu_quantum_2020,pinel_ultimate_2012,pinel_quantum_2013,shapiro_quantum_2009}, and our analysis
makes therefore heavy use of the QCRB for Gaussian states
\cite{sidhu_geometric_2020,safranek_estimation_2019,nichols_multiparameter_2018,braun_quantumenhanced_2018,holevo_statistical_1973,ragy_compatibility_2016}.

{The rest of the paper is organized as follows. In Section \ref{theory}, we describe the state for the $n$ mode interferometer for general sources on the source plane using the scattering matrix formalism. Later, we present the general formula of the POVM for the QFI based on the state of the $n$ mode interferometer. In Section \ref{result}, first, we discuss the QFI for the parameters of a single uniform circular source for both a single receiver and two receivers. Second, we discuss the spatial resolution of two strong point sources with the same and different temperatures for a two-mode interferometer. Third, we examine an array of receivers to increase the spatial resolution of a uniform circular source and two-point sources. We conclude in Section \ref{conclude}.}

\section{Theory}
\label{theory}
\subsection{Continuous Vector Potential and Interaction with Classical Current Sources}The operators for the quantized vector potential $\mathbf{A}(\mathbf{r}, t)$ can be written in continuous form. The operator for the vector potential in the Coulomb gauge
reads as \cite{blow_continuum_1990,mandel_optical_1996}

\begin{equation}
\begin{split}
\hat{\mathbf{A}}(\mathbf{r}, t)&=\int d^3 {k}\;\left(\frac{\hbar}{16 \pi^{3} \varepsilon_{0} c|\mathbf{k}|}\right)^{1 / 2}\\&\times\sum_{\sigma=1,2} \boldsymbol{\varepsilon}(\mathbf{k}, \sigma) \hat{a}(\mathbf{k}, \sigma) \exp (-i c|\mathbf{k}| t+i \mathbf{k} \cdot \mathbf{r}) +h.c.
\end{split}
\end{equation}
where, $  \hat{a}(\mathbf{k}, \sigma) $ are the continuous mode operators with {$[\hat{a}\left(\mathbf{k},\sigma\right),\hat{a}^\dagger\left(\mathbf{k},\sigma\right)] = \delta(\mathbf{k}-\mathbf{k}')\delta_{\sigma\sigma'}$, and $\boldsymbol{\varepsilon}(\mathbf{k}, \sigma)$ are the directions of the polarizations with index $\sigma \in 1,2$, which are always perpendicular to wave vector $\mathbf{k}$.} Mode functions are plane waves and parametrized by $\mathbf{k}$ and $\sigma$.
The interaction Hamiltonian for the classical current distribution of the sources $\mathbf{j}(\mathbf{r}, t)$ with electromagnetic waves in free space is given by \cite{glauber_coherent_1963, braun_fouriercorrelation_2018,mandel_optical_1996,scully_quantum_1999}
\begin{equation}
H_{I}(t)=-\int d^{3} {r}\; \mathbf{j}(\mathbf{r}, t) \cdot \hat{\mathbf{A}}(\mathbf{r}, t).
\end{equation}
In the interaction picture, using the Schrödinger equation the state of the electromagnetic field at time $t$ can be obtained from the one at $t_0$ as \cite{scully_quantum_1999,glauber_coherent_1963,mandel_optical_1996,loudon_quantum_1974}
\begin{equation}
|\psi(t)\rangle=U\left(t, t_{0}\right)\left|\psi\left(t_{0}\right)\right\rangle,
\end{equation}
where the $ U\left(t, t_{0}\right) $ is given by
\begin{equation}
	\small
	\begin{split}
		U\left(t, t_{0}\right) &=\exp \left(\frac{i}{\hbar} \int_{t_{0}}^{t} d t^{\prime} \int d^{3} {r}\; \mathbf{j}\left(\mathbf{r}, t^{\prime}\right) \cdot \hat{\mathbf{A}}\left(\mathbf{r}, t^{\prime}\right)+i\varphi\left(t, t_{0}\right)\right).
	\end{split}
\end{equation}
{The phase $\varphi\left(t, t_{0}\right)$ is a real 
number, which arises from the classical interaction between the currents. It is independent of the state on which the propagator acts, and cancels in the calculation of equal time matrix elements.} Since the current density commutes with the vector potential, one can write the time evolution in the form of a displacement operator, which is given by
\begin{equation}
\begin{split}
D(\{\alpha(\mathbf{k},\sigma)\})&=\exp \left[\sum_{\sigma}  \int d^3 {k}\;[\alpha(\mathbf{k},\sigma) \hat{a}^\dagger (\mathbf{k},\sigma)\right.\\&\left.-\alpha^*(\mathbf{k},\sigma) \hat{a} (\mathbf{k},\sigma) ] \right],
\end{split}
\end{equation}
where $\alpha(\mathbf{k},\sigma) $ can be found as
\begin{equation}
\begin{split}
\alpha(\mathbf{k},\sigma)&= \frac{i}{\hbar }\left(\frac{\hbar}{16 \pi^{3} \varepsilon_{0} c|\mathbf{k}|}\right)^{1 / 2}\int_{t_0}^{t} dt' \int d^3 {r}\;{\mathbf{j}}\left(\mathbf{r},t'\right) \cdot \boldsymbol{\varepsilon}(\mathbf{k}, \sigma) \\&\times \exp (i c|\mathbf{k}| t'-i \mathbf{k} \cdot \mathbf{r}).
\end{split}
\end{equation}

The $\alpha(\mathbf{k},\sigma)$ also depends on $t$ and $t_0$. We assume that for $t_0\rightarrow -\infty$ we have the vacuum state $\ket{\{0\}}$ for all modes. For a deterministic current density, $\ket{\psi(t)}$ is a tensor product of coherent states,
\begin{equation}
\ket{\psi(t)}=|\{\alpha(\mathbf{k},\sigma)\}\rangle = D(\{\alpha(\mathbf{k},\sigma)\})  |{\{0\}}\rangle.
\label{state}
\end{equation}
One can introduce the Fourier transform (FT) of the current densities and take the $t'$ integral immediately~\cite{braun_fouriercorrelation_2018}.  We introduce the Fourier decomposition of current density as
\begin{equation}
\mathbf{j}\left(\mathbf{r}, t'\right)=\frac{1}{\sqrt{2 \pi}} \int_{-\infty}^{\infty} d \tilde\omega  \tilde{\mathbf{j}}\left(\mathbf{r},\tilde \omega\right)\exp({i \tilde\omega t'}).
\end{equation}
Then we can write $\alpha(\mathbf{k},\sigma)$ in the following form

\begin{equation}
\begin{split}
\alpha(\mathbf{k},\sigma)=& \frac{i}{\hbar}\left(\frac{\hbar}{32 \pi^{4} \varepsilon_{0} c|\mathbf{k}|}\right)^{1 / 2}\\& \times\int_{-\infty}^{t} dt' \int d^3 {r}\;\int_{-\infty}^{\infty} d \tilde\omega  \tilde{\mathbf{j}}\left(\mathbf{r}^{\prime},\tilde \omega\right) \cdot \boldsymbol{\varepsilon}(\mathbf{k}, \sigma) \\&\times\exp (i c|\mathbf{k}| t'-i \mathbf{k} \cdot \mathbf{r})\exp({i \tilde\omega t'}).
\end{split}
\end{equation}
Taking the integral over $t'$ gives
\begin{equation}
\begin{split}
\alpha(\mathbf{k},\sigma)=& {-}\left(\frac{1}{32 \pi^{4} \varepsilon_{0} c\hbar|\mathbf{k}|}\right)^{1 / 2}\\&\times\int d^3 {r}\;\int_{-\infty}^{\infty} d \tilde\omega  \tilde{\mathbf{j}}\left(\mathbf{r},\tilde \omega\right) \cdot \boldsymbol{\varepsilon}(\mathbf{k}, \sigma) \exp (-i \mathbf{k} \cdot \mathbf{r}) \\&\times\frac{\exp({i( \tilde\omega + c|\mathbf{k}| )t})}{i\epsilon-c|\mathbf{k}|-\tilde\omega}.
\end{split}
\end{equation}

We introduced a shift in the denominator '$i\epsilon $'. that is necessary for the integral to converge at $t = -\infty$.

\subsection{The State Received by the Receivers} 
\begin{figure}[t!]
	\centering
	\includegraphics[width=0.99\linewidth]{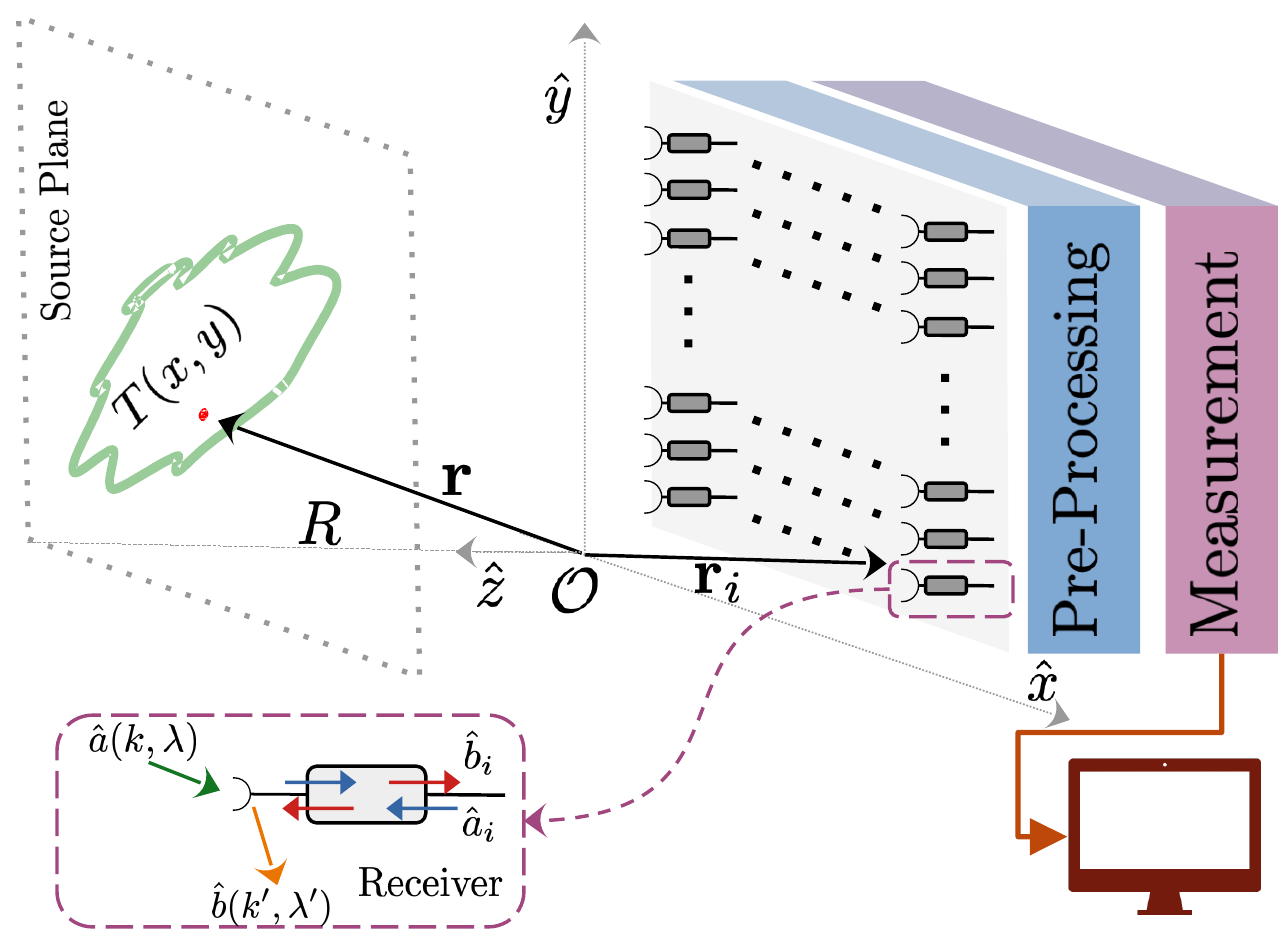}
	\caption{A interferometer, with $n$ receivers, separated by a distance $R$ from the source plane. The $T_{\mathrm{eff}}(x,y)$ is the position-dependent effective temperature in the source plane that one wants to measure. The field propagating from the source scatters each receiver on the interferometer. We consider the origin of the coordinate system on the detection plane as $\mathcal{O}$. All the components of the vectors are donated in the coordinate system $\mathcal{R}=(\mathcal{O},\hat{e}_x,\hat{e}_y,\hat{e}_z)$.}
	\label{interferometer}
\end{figure}

The electromagnetic field is received by an interferometer that has an array of receivers, localized at positions $\mathbf{r}_i$. {Each receiver is connected at its output to a waveguide that channels the received electromagnetic field radiation towards the measurement instruments. This output, possibly after filtering, is assumed to be single-mode with discrete annihilation operator $\hat{b}_i$, We call the modes received by the receivers ``spatial field modes'' since each mode $\hat b_i$ is specific to a location on the detection plane. A single-mode is assumed reflected from the measurement device with discrete annihilation operator $\hat{a}_{i}$}. On the antenna side, we represent incoming plane waves in the interferometer by $\hat{a}(\mathbf{k},\sigma)$ and scattered outgoing plane waves by $\hat{b}(\mathbf{k},\sigma)$. In Fig.~(\ref{interferometer}), we represent the current sources on the source plane and the $n$ mode array interferometer in the detection plane. One can use the scattering matrix formalism to find the relation between incoming and outgoing modes.

Furthermore, the modes $\hat{b}_i$ are separated by distances substantially larger than the central wavelength $\lambda $. And the collection area of each receiver $A_D$ is assumed to be $A_D \sim \lambda^2$, where $\lambda$ is central wavelength. These constraints make the modes for different receivers orthogonal and simplifies the form of the scattering matrix. A scattering matrix connects incoming and outgoing modes, and one can write it as \cite{zmuidzinas_cramer_2003, zmuidzinas_thermal_2003}

\begin{equation}
\mathcal{S} =\left[\begin{array}{ll}
\mathcal{S}^{(\text {scat })} & \mathcal{S}^{(\mathrm{trans})} \\
\mathcal{S}^{(\text {rec })} & \mathcal{S}^{(\text {refl })}
\end{array}\right].
\end{equation}

This matrix acts on the vector $\left[\hat{a}(\mathbf{k},\sigma), \hat{a}_i \right]^T$. The first block, $\mathcal{S}^{(\text {scat })}$, describes the scattering of incoming plane waves to outgoing plane waves from the interferometer. A receiver can receive or transmit the signal. The off-diagonal block $\mathcal{S}^{(\mathrm{rec})}$ describes the coupling of the incoming plane waves $ \hat a (\mathbf{k},\sigma)$ into the receiver modes $\hat b_i$, and $\mathcal{S}^{(\operatorname{trans})}$ describes scattering of reflected receiver modes $\hat a_i $ into outgoing plane waves $\hat b(\mathbf{k},\sigma)$. The matrix $\mathcal{S}^{(\text {refl })}$ represents the scattering (reflection) between the receivers, and will be neglected, $\mathcal{S}^{(\text {refl })} \sim 0 $.  One can also verify that if the receivers have only incoming and outgoing modes, the receiving and transmitting pattern of the receivers will be the same {$\mathcal{S}^{(\operatorname{trans})}(\mathbf{k},\sigma;j)= \mathcal{S}^{(\operatorname{rec})}(j;\mathbf{k},\sigma)$ and we can denote them as simply $\mathcal{S}_{j }(\mathbf{k},\sigma)$.} Formally, the input-output relations read
\begin{equation}
\begin{split}
\hat{b}(\mathbf{k},\sigma)=&\sum_{\sigma'}\int {d}^3 \mathbf{k}^{\prime} \mathcal{S}^{(\mathrm{scat})}\left(\mathbf{k}, \mathbf{k}^{\prime},\sigma,\sigma'\right) \hat{a}\left(\mathbf{k}^{\prime},\sigma'\right)\\&+\sum_{j } \mathcal{S}_{j }(\mathbf{k},\sigma) \hat{a}_{j},
\end{split}
\end{equation}
and,
\begin{equation}
\hat{b}_{i }=\sum_\sigma\int \mathrm{d}^3 \mathbf{k}\mathcal{S}_{i}\left(\mathbf{k},\sigma\right) \hat{a}\left(\mathbf{k},\sigma\right).
\end{equation}
For a lossless system we can assume that $S^\dagger S = I $. Then we can write $ \mathcal{S}^{(\mathrm{scat})}\left(\mathbf{k}, \mathbf{k}^{\prime},\sigma,\sigma'\right)   = (\mathcal{S}^{(\mathrm{scat})})^T\left(\mathbf{k}', \mathbf{k},\sigma',\sigma\right)$. The field operators $ \hat a (\mathbf{k},\sigma)$ from the state that we have for Eq.~(\ref{state}) can be replaced by the following relation for $n$ different receiver modes
\begin{equation}
	\begin{split}
		\hat{a}(\mathbf{k},\sigma) &= \sum_j^n {\mathcal{S}^*_{j}(\mathbf{k},\sigma)}\hat{b}_j \\&+ \sum_{\sigma'}\int d^3\mathbf{k'}{\mathcal{S}^{*(\mathrm{scat})}(\mathbf{k}',\sigma',\mathbf{k},\sigma)}\hat{b}(\mathbf{k}',\sigma').
	\end{split}
\label{eq1}
\end{equation}
{The interferometer does not have any access to modes $\hat{b}(\mathbf{k},\sigma)$.} For an array of receivers at positions $\mathbf{r}_i$ in the detection plane, each scattering function may be written as \cite{zmuidzinas_cramer_2003}
\begin{equation}
\begin{split}
 \mathcal{S}_{i}(\mathbf{k},\sigma)= e^{i\mathbf{k}\cdot \mathbf{r}_i-\omega t_i} \mathcal{S}(\mathbf{k},\sigma),
\end{split}
\end{equation} where $t_i$ is the time at which we consider the state of the $i$-th receiver. We can assume that for all receivers 
$t_i\equiv \bar t > t-|\mathbf{r}-\mathbf{r}_i|/c$ due to retardation, where according to \eqref{eq1}, $t$ is the last time the current densities to be sensed imprint their information the coherent state labels $\alpha(\bm k,\sigma)$. And $ \mathcal{S}(\mathbf{k},\sigma)$ is the function describes scattering to the central receiver. Further, the commutation relation of different receiver modes can be written as
\begin{equation}
[\hat{b}_i,\hat{b}_j^\dagger] = \sum_\sigma\int \mathrm{d}^3 \mathbf{k}\mathcal{S}_{i}\left(\mathbf{k},\sigma\right) \mathcal{S}_{j}^{*}\left(\mathbf{k},\sigma\right) \approx \delta_{ij}.
\label{eq16}
\end{equation}
{where we have used the canonic commutation relation of $\hat{a}\left(\mathbf{k},\sigma\right)$ and we assumed that $|\mathcal{S}(\mathbf{k},\sigma)|$ varies slowly compared to the oscillations of the exponential factor $\exp(i\mathbf{k}\cdot(\mathbf{r}_i-\mathbf{r}_j))$ for $i\neq j$. Since $\hat{b}_i $ commutes with $\hat{b}(\mathbf{k},\sigma)$, using Eq.~(\ref{eq1}) we can write the coherent state in Eq.~\eqref{state} as}
\small
\begin{equation}
\begin{split}
	\ket{\psi(t)} =  D(\{\beta_i\})   D(\{\beta(\mathbf{k},\sigma)\})  \ket{\{0\}}.
\end{split}
\end{equation}
\normalsize
After, tracing out the modes $\hat{b}(\mathbf{k},\sigma)$ we have a coherent state
\begin{equation}
\begin{split}
\rho' =  \ket{\{\beta_i\}} \bra{\{\beta_i\}},
\end{split}
\label{state2}
\end{equation}
and the displacement operator can be written in the form
\begin{equation}
\begin{split}
D(\{\beta_i\}) =   \bigotimes_i^n\mathrm{exp} \left[\beta_i\hat{b}_i^\dagger-\beta_i^*\hat{b}_i\right]
\end{split}
\end{equation}
where
\begin{equation}
\begin{split}
\beta_i &=\sum_{\sigma}  \int d^3 {k}\; S_i(\mathbf{k},\sigma)\alpha(\mathbf{k},\sigma)\,.
\end{split}
\end{equation}
$\mathcal{S}_{i}(\mathbf{k},\sigma)$ depends on the type of receivers. Let us assume that each receiver is characterized by a filter function $w(\omega)$ with central frequency $\omega_0$ and bandwidth $B\ll \omega_0$,
\begin{equation}
w(\omega)=\left\{\begin{array}{ll}
1 & \text { for } \omega_{0}-B/ 2 \leqslant \omega \leqslant \omega_{0}+B / 2 \\
0 & \text { elsewhere }
\end{array}\right. .
\label{eq:filter}
\end{equation}
For simplicity we assume $S(\mathbf{k},\sigma)\propto \sqrt{\omega}w(\omega) \boldsymbol{\varepsilon}(\mathbf{k}, \sigma)\cdot\hat{u}$, and normalized according to Eq.~(\ref{eq16}) as
\begin{equation}
	S(\mathbf{k},\sigma)= \left(\frac{3c^3 \omega }{8\pi\omega_0 ^3B}\right)^{1/2}w(\omega) \boldsymbol{\varepsilon}(\mathbf{k}, \sigma)\cdot\hat{u}
\end{equation}
where $\omega =c|\mathbf{k}|$ and $\hat{u}$ is the unit polarization direction of the corresponding receiver mode. Then we have
\begin{equation}
\begin{split}
\beta_i =& { -\left(\frac{3c^3}{2^8\hbar \varepsilon_0 \pi^{5} \omega_0^3 B}\right)^{1 / 2}} \int d^3 {r}\;\int_{-\infty}^{\infty} d \tilde\omega  \\& \times\sum_{\sigma}  \int d^3 {k}\;{w(\omega)}\tilde{\mathbf{j}}\left(\mathbf{r},\tilde \omega\right) \cdot \boldsymbol{\varepsilon}(\mathbf{k}, \sigma)\boldsymbol{\varepsilon}(\mathbf{k}, \sigma)\cdot\hat{u} \\&\times\frac{e^{i (\tilde\omega t+\omega t - \omega \bar t)} e^{ -i \mathbf{k} \cdot (\mathbf{r}-\mathbf{r}_i)}}{i\epsilon-c|\mathbf{k}|-\tilde\omega}.
\end{split}
\end{equation}
To take the integral over $d^3{k}$ we align the $k_z$-axis with the vector $\left(\mathbf{r}-\mathbf{r}_{i}\right)$.
In spherical coordinates in $k$-space we have $d^3{k}=\omega^2/c^3 d\omega d\Omega$, where $\omega = |\mathbf{k}|c$ and $\mathbf{k}= (\omega/c )\hat{\mathbf{n}}(\Omega)$ with $\hat{\mathbf{n}}(\Omega)=(\sin\theta\cos \phi, \sin \theta \sin \phi, \cos \theta) $. The two polarizations can be written in the form $ \boldsymbol{\varepsilon}(\Omega,1)=(\sin \phi,-\cos \phi, 0) $ and $\boldsymbol{\varepsilon}(\Omega,2)=(\cos \theta \cos \phi, \cos \theta \sin \phi,-\sin \theta)$. Taking the integral over $\Omega$, summing over two polarizations, and using one of the approximations of the far field limit $\omega|\mathbf{r}-\mathbf{r}_i|/c\gg 1$, 
gives {
\begin{equation}
\begin{split}
\beta_i =& {i} \left(\frac{3c\mu_0}{64 \hbar \pi^3 \omega_0^3 B}\right)^{1 / 2}
 \int d^{3} {r} \int_{-\infty}^{\infty} d\tilde{\omega}\int_0^\infty d\omega {w(\omega)}{\omega} \\&\times\tilde{\mathbf{j}}_t\left(\mathbf{r},\tilde{\omega}\right) \cdot\hat{u} \frac{ e^{i\omega|\mathbf{r}-\mathbf{r}_i|/c}-e^{-i\omega|\mathbf{r}-\mathbf{r}_i|/c}}{|\mathbf{r}-\mathbf{r}_i|}\frac{e^{i (\tilde\omega t+\omega t - \omega \bar t)}}{i\epsilon-\omega-\tilde\omega},
\end{split}
\end{equation}
}
{where $\tilde{\mathbf{j}}_t\left(\mathbf{r},\tilde{\omega}\right)  $ is the locally transverse component of the current density defined as $\tilde{\mathbf{j}}_{t}=\tilde{\mathbf{j}}-(\tilde{\mathbf{j}} \cdot \hat{e}_{\mathbf{r}})\hat{e}_{\mathbf{r}}$ with unit vector $ \hat{e}_{\mathbf{r}} = (\mathbf{r}-\mathbf{r}_i)/|\mathbf{r}-\mathbf{r}_i|. $ Since $R \gg |\mathbf{r}_i| $, we have $ \hat{e}_{\mathbf{r}}  \approx \mathbf{r}/|\mathbf{r}|$, with corrections modifying only slightly the prefactors, not the phases. One can extend the lower bound of the integration range of the ${\omega}$ integral to $-\infty$ using the definition of $w(\omega)$, and evaluate the $\omega$ integral with the help of the law of residues. Since $-\bar t +t-|\mathbf{r}-\mathbf{r}_i|/c <0 $, the pole at $\omega = -\tilde\omega + i\epsilon $ contributes to the term $\exp{(i\omega|\mathbf{r}-\mathbf{r}_i|/c)}$. For $\exp{(-i\omega|\mathbf{r}-\mathbf{r}_i|/c)}$ the contour must be closed in the lower half plane and there is no pole to contribute. In the end one should send $\epsilon \rightarrow 0$. Then $ \beta_i$ simplifies to
\begin{equation}
\begin{split}
 \beta_i =&-\left(\frac{3c\mu_0}{16\pi\hbar \omega_0^3 B}\right)^{1 / 2}
 \int_{-\infty}^\infty d\omega {w(-\omega)}{\omega}\int d^{3} {r} \\&\times\tilde{\mathbf{j}}_t\left(\mathbf{r},{\omega}\right) \cdot\hat{u} \frac{e^{-i\omega(\bar t-|\mathbf{r}-\mathbf{r}_i|/c})}{|\mathbf{r}-\mathbf{r}_i|}.
\end{split}
\end{equation}
where we drop the "$\sim$" from $\tilde\omega$.} The state in Eq.~(\ref{state}) is written for a deterministic current density distribution. In reality, these current densities fluctuate. Before we move forward, we describe the properties of these current density distribution. We assume that it is a complex symmetric Gaussian process with current densities uncorrelated  in positions, directions and frequencies \cite{braun_fouriercorrelation_2018,kubo_fluctuationdissipation_1966,savasta_light_2002}, 
\begin{equation}
\small
\begin{split}
&\left\langle\tilde{j}_{i}\left(\mathbf{r}, \omega\right) \tilde{j}_{j}^{*}\left(\mathbf{r}^{\prime}, \omega'\right)\right\rangle =\frac{l_c^3}{\tau_c}  \delta_{ij}\delta(\omega-\omega')\delta(\mathbf{r}-\mathbf{r}')\braket{|\tilde{{j}}_{i}(\mathbf{r},\omega)|^2}, \\& \left\langle\tilde{j}_{i}\left(\mathbf{r}, \omega\right) \tilde{j}_{j}\left(\mathbf{r}^{\prime}, \omega'\right)\right\rangle =0,  \quad  \left\langle\tilde{j}^{*}_{i}\left(\mathbf{r}, \omega\right) \tilde{j}_{j}^{*}\left(\mathbf{r}^{\prime}, \omega'\right)\right\rangle =0\,.
\end{split}
\label{current}
\end{equation}
The length scale $l_c$ and time scale $\tau_c$ are introduced for dimensional grounds. We choose the unit polarization vector of the receiver $\hat{u}$ as one of basis vectors of the coordinate system 
$\mathcal{R}$ parallel to   the detection plane, $\hat{u} 
=\hat{e}_i $ where $i\in \{x,y\} $. Then, we can write $\langle{\tilde{\mathbf{j}}_t(\mathbf{r}, \omega) \cdot\hat{u} \tilde{\mathbf{j}}^*_t(\mathbf{r}, \omega) \cdot\hat{u}}\rangle = \langle{|\tilde{{j}}_{t,i}(\mathbf{r}, \omega) |^2}\rangle $. Using Eq.~(\ref{state2}) and introducing the distribution of the current density 
$P({\tilde j }({r},\omega) )$, the state for the interferometer $\rho_{\mathrm{int}}$ with $n$ receivers can be written as
\begin{equation}
\rho_{\mathrm{int}} =  \int d^2{\tilde j} P({\tilde j}(\mathbf{r},\omega) )|\{\beta_i\}\rangle \langle\{\beta_i\}|.
\end{equation}
The integral is over the complex ${\tilde j} $ plane. Gaussian states are completely characterized by their mean displacement $\Gamma_{i}=\operatorname{Tr}\left[\rho \mathbf{b}_{i}\right]$ and covariance matrix with elements $\Sigma_{ij}=\frac{1}{2} \operatorname{Tr}\left[\rho\left(\tilde{\mathbf{b}}_{i} \tilde{\mathbf{b}}_{j}+\tilde{\mathbf{b}}_{j} \tilde{\mathbf{b}}_{i}\right)\right],$ where $\mathbf{b}=\left[b_{1}, b_{1}^{\dagger}, b_{2}, b_{2}^{\dagger}, ...b_n,b_n^\dagger\right]$ and $\quad \tilde{\mathbf{b}}_{i}=\mathbf{b}_{i}-\Gamma_{i}$ ~\cite{braun_precision_2014,adesso_continuous_2014,gao_bounds_2014,olivares_quantum_2012,pinel_ultimate_2012,weedbrook_gaussian_2012}. The mean displacement for our state is zero $\Gamma_i = 0$ considering Eq.~(\ref{current}). To find the elements of the covariance matrix, we need to calculate $ \braket{b^\dagger_i b_j}$. The integral over $\omega $ can be taken using the filter function of bandwidth $B$. With this we find
\begin{equation}
\begin{split}
\braket{b^\dagger_i b_j} &=K\int d^{3} {r}\;\frac{\braket{|\tilde{{j}}_{t,i}\left(\mathbf{r},{\omega}\right)|^2}e^{i\omega_0(|\mathbf{r}-\mathbf{r}_j|-|\mathbf{r}-\mathbf{r}_i|)/c}}{|\mathbf{r}-\mathbf{r}_i||\mathbf{r}-\mathbf{r}_j|}\\&\times\mathrm{sinc}\left[\frac{B}{2c}(|\mathbf{r}-\mathbf{r}_j|-|\mathbf{r}-\mathbf{r}_i|)\right]
\label{eq26}
\end{split}
\end{equation}
where, $K = 3c\mu_0l_c^3 /(16\pi\hbar\omega_0\tau_c)$ and $\mathrm{sinc}[x] \equiv \sin x/x$. For a very narrow bandwidth $\mathrm{sinc}[...]  \approx 1$. Then Eq.~(\ref{eq26}) for $i = j $ becomes
\begin{equation}
\begin{split}
\bar n &=K\int d^{3} {r}\;\frac{\braket{|\tilde{{j}}_{t,i}\left(\mathbf{r},{\omega}\right)|^2}}{|\mathbf{r}-\mathbf{r}_i|^2}\,.
\end{split}
\label{eq27}
\end{equation}
where we defined $\bar n \equiv \braket{b^\dagger_i b_i}$ without any index, since the mean photon number is the same for all interferometer modes in the far-field approximation,
and for $i \neq j $ it becomes 
\begin{equation}
	\begin{split}
		\xi_{ij} &=K\int d^{3} {r}\;\frac{\braket{|\tilde{{j}}_{t,i}\left(\mathbf{r},{\omega}\right)|^2}e^{i\omega_0(|\mathbf{r}-\mathbf{r}_j|-|\mathbf{r}-\mathbf{r}_i|)/c}}{|\mathbf{r}-\mathbf{r}_i||\mathbf{r}-\mathbf{r}_j|}\,.
	\end{split}
	\label{eq28}
\end{equation}
with $\xi_{ij}\equiv \braket{b^\dagger_i b_j}$. The integral over Earth's surface is parametrized by $\mathbf{r}= (x,y,R)$ with respect to the coordinate system of the detection plane. Further, we write $|\mathbf{r}-\mathbf{r}_j|-|\mathbf{r}-\mathbf{r}_i| \approx \Delta\mathbf{r}_{ij}\cdot \mathbf{r}/|\mathbf{r|}$ for $|\Delta\mathbf{r}_{ij}|\ll R$, where $\Delta\mathbf{r}_{ij}= \mathbf{r}_j-\mathbf{r}_i$ connects two different receiver modes. In the denominator, we approximate $|\mathbf{r}-\mathbf{r}_i| \approx R/\cos \tilde\theta(x,y) $ with $\tilde\theta(x,y)$ the polar angle the angle between the $z$-axis and the vector $(x,y,R)$. One can relate the average amplitude of current density to brightness temperature $T_{\mathrm{B}}(x,y)$ by $\braket{|\tilde{{j}}_{t,i}\left(\mathbf{r},{\omega}\right)|^2} = K_1 T_{\mathrm{B}}(x,y)\cos\tilde \theta(x,y)\delta(z-R)$ with a constant defined as  $K_1 = 32 \tau_c k_B /(3 l_c^3 \mu _0 c)$ (See Appendix \ref{appendixAnew}). We define the effective temperature as $T_{\mathrm{eff}}(x,y)\equiv T_{\mathrm{B}}(x,y) \cos^3\tilde \theta(x,y)$ and a new constant $
\kappa =  K_1K \equiv {2k_B}/{(\pi \hbar \omega_0)} $ where $\kappa$
has the dimension of inverse temperature with SI-units
"$[1/\mathrm{K}]$". 
Then we can simplify Eq.~(\ref{eq27}) for $i=j$ as
\begin{equation}
\begin{split}
\bar{n} &=\frac{\kappa}{R^2}\int dxdy\; T_{\mathrm{eff}}(x,y),
\end{split}
\label{eqn}
\end{equation}
 and for $i\neq j$ as
\begin{equation}
\begin{split}
\xi_{ij}&=\frac{\kappa}{R^2}\int dxdy\;T_{\mathrm{eff}}(x,y)e^{2\pi i\left(v_x^{ij}x +v_y^{ij}y \right)},
\end{split}
\label{eqxi}
\end{equation}
where
\begin{equation}
\begin{split}
    v_y^{ij} = \frac{\Delta x_{ij} }{\lambda R},\qquad v_x^{ij} = \frac{\Delta y_{ij} }{\lambda R}.
\end{split}
\end{equation}
We used $\omega_0/c = 2\pi /\lambda$. These two equations suffice to determine the covariance matrix elements of the Gaussian states for the general interferometer with an array of receivers. All spatial field modes received by the interferometer undergo a preprocessing before measurement. This processing can be understood as a linear combination of all spatial modes in such a way to achieve the optimal POVM for the best estimation of the parameter we are interested in. See section \ref{sec2c}. We use the values of the SMOS for the rest of the paper which leads to $\kappa \sim 9.4 \:[1/\mathrm{K}]$.

\subsection{Quantum Cram\'er-Rao Bound}
\label{sec2c}
{
A lower bound of an unbiased estimator of a deterministic parameter is given by the Cram\'er-Rao (CR) bound, which states that the variance of any such estimator is equal or greater than the inverse of the Fisher information. The quantum analog of the Cram\'er-Rao bound is the quantum Cram\'er-Rao bound (QCRB), given by the inverse of the quantum Fisher information (QFI).} The significance of the QCRB lies in the fact that in the case of a single parameter to be estimated the bound can in principle be saturated in the limit of infinitely many measurements when chosing the optimal quantum measurement and maximum-likelihood estimation.  Let us consider a quantum state $\rho_{\boldsymbol{\mu}}$ that depends on a vector of $l$ parameters, $ \boldsymbol{\mu} = (\mu_1,\mu_2,...,\mu_l)^T$. One can generalize the single-parameter quantum Cram\'er-Rao bound (QCRB) ~\cite{helstrom_quantum_1969,helstrom_detection_1967} to the multiparameter QCRB ~\cite{szczykulska_multiparameter_2016} given {for a single measurement} by
\begin{equation}
	\operatorname{Cov}(\tilde{\boldsymbol{\mu}}) \geqslant \mathscr{F}(\boldsymbol{\mu})^{-1}, \quad \mathscr{F}_{{i j}}(\boldsymbol{\mu})=\frac{1}{2} \operatorname{tr}\left(\rho_{\boldsymbol{\mu}}\left\{\mathscr{L}_{i}, \mathscr{L}_{j}\right\}\right),\label{multi}
\end{equation}
where $\operatorname{Cov}(\tilde{\boldsymbol{\mu}}) $ is a covariance matrix for the locally unbiased estimator $ \tilde{\boldsymbol{\mu}}(x)$ \cite{ragy_compatibility_2016,sidhu_geometric_2020}, the $\{\cdot,\cdot \}$ means the anti-commutator, and $\mathscr{L}_{i}$ is the symmetric logarithmic derivative (SLD) related to parameter $i$, which is defined similarly to the single-parameter case, $\frac{1}{2}\left(\mathscr{L}_{i} \rho_{\boldsymbol{\mu}}+\rho_{\boldsymbol{\mu}} \mathscr{L}_{i}\right)=\partial_{{i}} \rho_{\boldsymbol{\mu}} .$  {Contrary to the single parameter case, the multiparameter QCRB can in general not be saturated, but gives a useful lower bound nevertheless.}

 The SLD and the elements of QFI matrix are given in Ref.~\cite{gao_bounds_2014} for any Gaussian state. The SLD can be written as
 \begin{equation}
 \mathscr{L}_{i}=\frac{1}{2} \mathfrak{M}_{\alpha \beta, \gamma \delta}^{-1}\left(\partial_{i} \Sigma^{\gamma \delta}\right)\left(\mathbf{b}_{\alpha} \mathbf{b}_{\beta}-\Sigma^{\alpha \beta}\right),
 \end{equation}
 where the summation convention is used. In our case, the mean displacement of Gaussian state is zero. Thus, we can simplify further the elements of the QFI matrix in \cite{gao_bounds_2014} to
\begin{equation}
\mathscr{F}_{i j}=\frac{1}{2} \mathfrak{M}_{\alpha \beta, \gamma \delta}^{-1} \partial_{j} \Sigma^{\alpha \beta} \partial_{i} \Sigma^{\gamma \delta},
\end{equation}
where
\begin{equation}
\mathfrak{M} \equiv \Sigma \otimes \Sigma+\frac{1}{4}\Omega \otimes \Omega ,
\end{equation}
and $\Omega=\bigoplus_{k=1}^{n} i \sigma_{y}$.
Using the properties of the Gaussian state (circularly symmetric and with zero mean) we can write the SLD for $n$ mode interferometers as
\begin{equation}
    \begin{split}
      \mathscr{L}_{i} = \sum_j^n g^{j}_i \hat{b}^\dagger _j \hat{b}_j + \sum _{j<k}^n (g^{jk}_i\hat{b}^\dagger _j \hat{b}_k +  (g^{jk}_{i})^*\hat{b}^\dagger _k \hat{b}_j) +\mathrm{C},
    \end{split}
	\label{eq:Sld}
\end{equation}
where C is a constant term. In the single parameter case, the optimal POVM is the set of
projectors onto eigenstates of $\mathscr{L}_i$.  It allows one to 
saturate the QCRB in the limit of infinitely many measurements and 
maximum likelihood estimation \cite{helstrom_detection_1967,braunstein_statistical_1994,paris_quantum_2009}. 
In the multiparameter case, \eqref{multi} can in general not be saturated.
For the diagonalization of the SLD, the constant C is not important and we can drop it from the beginning. {We construct a Hermitian matrix $\mathbf{M}_i$
\begin{equation}
\mathbf{M}_i=\left[\begin{array}{cccc}
g^{1}_i & g^{12}_i  & ... & g^{1n}_i  \\
(g^{12}_{i})^* & g^{2}_i & ... &g^{2n}_i  \\
... & ... & ... & ... \\
(g^{1n}_{i})^* & (g^{2n}_{i})^* & ... & g^{n}_i
\end{array}\right],
\end{equation}
where the diagonal elements are real-valued functions which can be defined as $g^{j}_i =\mathfrak{M}_{\alpha \beta, \gamma \delta}^{-1}\left(\partial_{i} \Sigma^{\gamma \delta}\right) $ with $\alpha = 2j$ and $\beta = 2j-1$. The off-diagonal elements are complex-valued functions which are defined as $g^{jk}_i =\mathfrak{M}_{\alpha \beta, \gamma \delta}^{-1}\left(\partial_{i} \Sigma^{\gamma \delta}\right) $ }with $\alpha = 2j$ and $\beta = 2k-1$ and $k>j$.
Further, we can define a new set of operators $\bar{\mathbf{b}}^\dagger \equiv {\left[\hat b_1^\dagger, \hat b_2^\dagger, ..., \hat b_n^\dagger\right]}$ and $\bar{\mathbf{b}} \equiv {\left[ \hat b_1, \hat b_2, ..., \hat b_n\right]^T}$. {Then the SLD becomes
\begin{equation}
    \begin{split}
      \mathscr{L}_{i} = \bar{\mathbf{b}}^\dagger \mathbf{M}_i \bar{\mathbf{b}}.
    \end{split}
\end{equation}
Since $\mathbf{M}_i$ is a Hermitian matrix, it can always be unitarily
diagonalized by $\mathbf{M}_i =\mathbf{V}^\dagger_i \mathbf{D}_i
\mathbf{V}_i $ with $\mathbf{V}_i^\dagger \mathbf{V}_i = I$.} A new
set of operators can be defined as $\bar{\mathbf{d}}_i^\dagger =
\bar{\mathbf{b}}^\dagger \mathbf{V}_i^\dagger$ where
$\bar{\mathbf{d}}_i^\dagger = {\left[\hat d_{i1}^\dagger, \hat
    d_{i2}^\dagger, ..., \hat d_{in}^\dagger\right]}$. The optimal
POVM for the single parameter case ($i=1$, which we drop in the
following) can be found as a set of projectors in the Fock basis $\{\ket{m_1,m_2,...,m_n}\bra{m_1,m_2,...,m_n}\}_{\{m_1,m_2...m_n\}}$ of the $\hat d_l $ with $\hat d^\dagger_l \hat d_l\ket{m_1,m_2,...,m_n} = m_l\ket{m_1,m_2,...,m_n}$, where $l\in \{1,...,n\}$. The $\hat{d}_l$ will be called "detection modes".

\section{Results}
\label{result}
\subsection{The Single Receiver}
In this section, we consider the case of the simplest estimation of the parameters of the sources with a single receiver with mode $\hat{b}$. Then the covariance matrix for the state can be written as
\begin{equation}
\begin{aligned}
\Sigma &=\left[\begin{array}{cccc}
0 & \chi \\
\chi & 0
\end{array}\right].
\end{aligned}
\end{equation}
The QFI matrix elements for single mode can be found as
\begin{equation}
	\begin{split}
		\mathscr{F}_{ij} = \frac{4\partial_{i} \chi\partial_{j}\chi}{4\chi^2-1},
	\end{split}
\end{equation}
and, up to the irrelavant constant, the SLD becomes, 
\begin{equation}
	\begin{split}
		\mathscr{L}_{i} = \frac{4\partial_{i} \chi}{4\chi^2-1} \hat{b}^\dagger b.
	\end{split}
\end{equation}
Since the SLD is already diagonal in the basis of $\hat b^\dagger \hat b$, the detection mode can be considered as $\hat b$. We write the POVM obtained from the SLD as a set of projectors in the Fock basis $\{\ket{m}\bra{m}\}_{\{m\}}$ which is the eigenbasis of $\hat b^\dagger \hat b $, $ \hat b^\dagger\hat b\ket{m}\bra{m}= m\ket{m}\bra{m}$. To compare, we consider the POVM from heterodyne detection. The heterodyne detection uses a classical local oscillator to make a measurement locally on the basis of coherent states. For a single mode, its POVM elements can be written as $E(\nu)=\ket{\nu}\bra{\nu}/\pi$ where $\ket{\nu}$ is coherent state and $\int d^2 \nu E(\nu) = \mathds{1}$. The probability that $E(\nu)$ triggers reads
\begin{equation}
	\begin{split}
		P(\nu|\mu_i) = \frac{1}{\pi (1+\bar n)}\exp \left[-\frac{|\nu|^2}{(1+\bar n)} \right]\,,
	\end{split}
      \end{equation}
      with $\bar n$ given by eq.\eqref{eqn}. 
The classical Fisher information (CFI) for parameter $\mu_i$ can be written as
\begin{equation}
	\begin{split}
		F_{i}&= \int d^2\nu \frac{1}{P(\nu|\mu_i)}\left(\frac{\partial P(\nu|\mu_i)}{\partial \mu_i}\right)^2.
	\end{split}
\end{equation}

\begin{figure}[t!]
	\centering
	\subfloat[][]{%
		\includegraphics[width=0.49\linewidth]{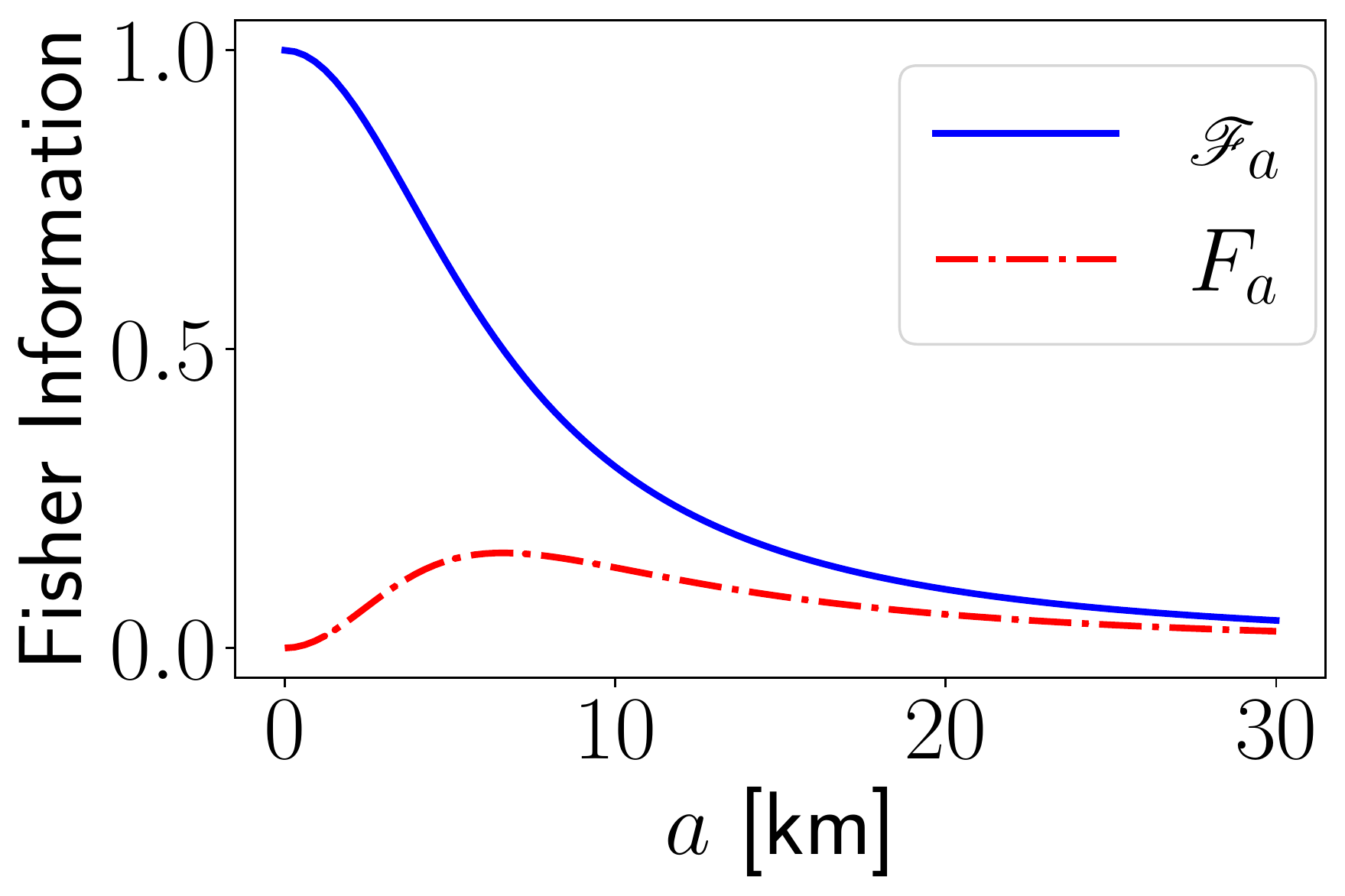}
		\label{figa:single}
	}
	\subfloat[][]{%
		\includegraphics[width=0.49\linewidth]{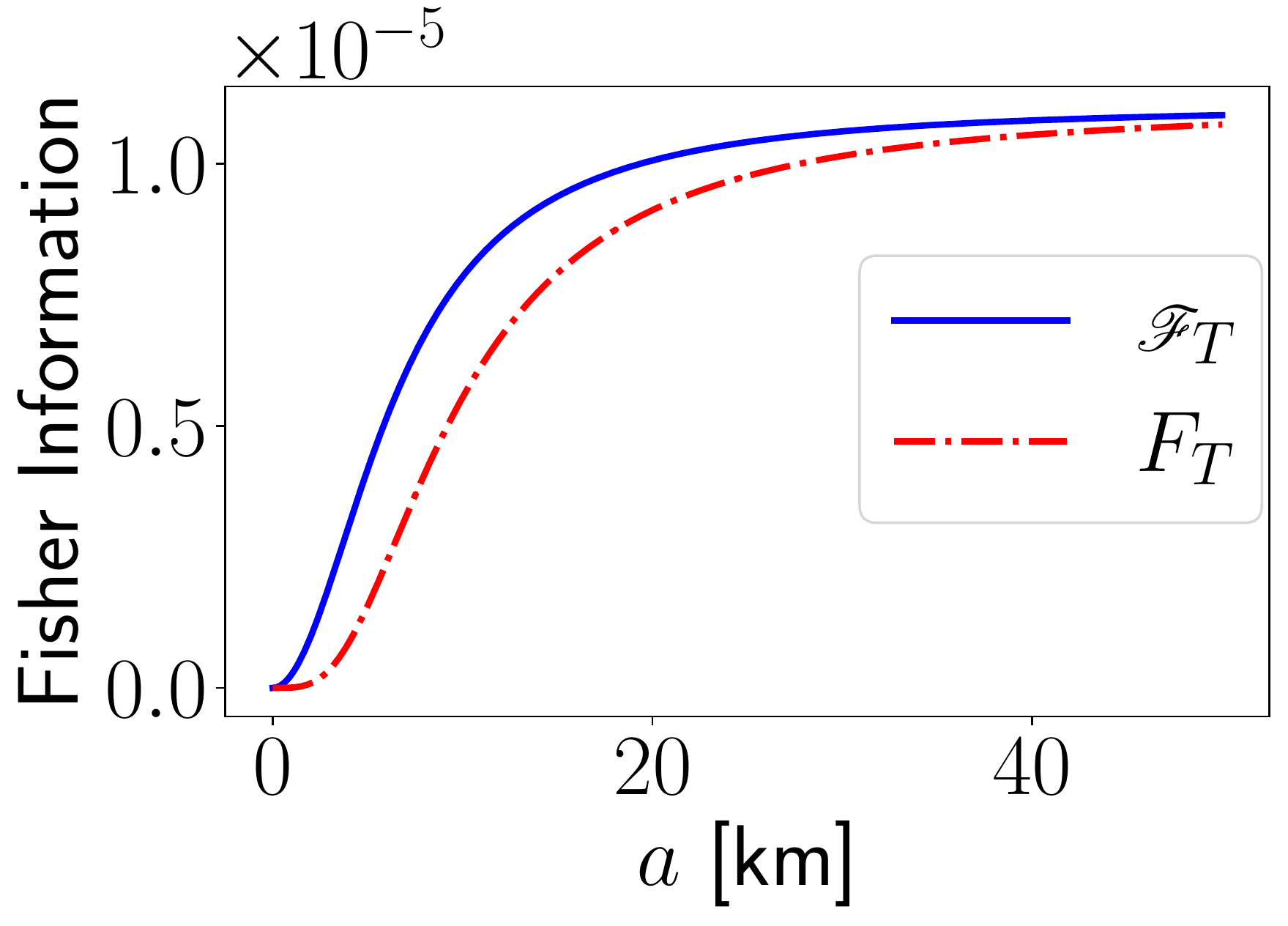}
		\label{figT:single}
	}
	\caption{a) The Fisher information (dimensionless) concerning source size. The solid blue line shows the QFI for estimating the source size. The red dashed line shows the CFI that one gets from the heterodyne measurement. Both results are scaled with the $4\pi T\kappa /R^2$, considering $T=300$ [K]. b) The Fisher information concerning the temperature. The solid blue line shows the QFI for estimating the temperature. The red dashed line shows the CFI that one gets from the heterodyne measurement. Both results are in units of [$\mathrm{1/K^2}$], and we consider $T=300$ [K].}
	\label{figsingle}
\end{figure}

\textit{The Resolution of a Uniform Circular Source:} Consider a
source defined as a circular disk with radius $a$ and with uniform
temperature $T$ located under the interferometer at a distance $R$,
($\mathbf{r}=(0, 0, R)$). We assume $a\ll R$. Then only small angles are involved and one can set $\cos^3\Theta(x,y)
\approx 1$. This corresponds to one of the approximations
characteristic of the far field regime
\cite{goodman_statistical_1985}.   
\begin{equation}
    T_{\mathrm{eff}}(x,y) = T \operatorname{circ}(x,y),
	\label{Eqtemp}
\end{equation}
where the symbol $\mathrm{circ}(\cdot)$ stands for the circular function, defined as
\begin{equation}
\operatorname{circ}(x,y) \triangleq\left\{\begin{array}{ll}
1 & \sqrt{x^2+y^2} \leq a \\
0 &  \sqrt{x^2+y^2}>a
\end{array}\right.
\end{equation}
\begin{equation}
\begin{split}
\bar{n} &=\frac{\kappa T}{R^2}\int dxdy \operatorname{circ}(x,y) \\&= \frac{\pi a^2 \kappa T}{R^2},
\end{split}
\end{equation}
and $\chi$ becomes $\chi = {1}/{2}+ \bar{n}$. We may consider $a$ and $T$ as the parameter that we want to estimate. The QFI for estimating $a$ becomes
\begin{equation}
	\begin{split}
		\mathscr{F}_a = \frac{4\pi T\kappa}{R^2+a^2\pi T \kappa }.
	\end{split}
	\label{Eqasingle}
\end{equation}
Then we can write the SLD for estimating the $a$ ignoring the constant term as {
\begin{equation}
	\begin{split}
		\mathscr{L}_{a} = \frac{2R^2}{aR^2 + a^3\pi T\kappa}\hat b^\dagger \hat b.
	\end{split}
\end{equation}
}The optimal POVM is a set of projectors in the Fock basis $\{\ket{m}\bra{m}\}_{\{m\}}$. The CFI of the heterodyne detection becomes
\begin{equation}
	\begin{split}
		{F}_a = \frac{4a^2 \pi^2  T^2 \kappa^2}{(R^2+a^2\pi T \kappa)^2  }.
	\end{split}
\end{equation}
In Fig.~(\ref{figsingle}), we compare the QFI with the CFI of
heterodyne detection. As one can see, for small source sizes, the
Fisher information from heterodyne measurement vanishes. However, the
QFI tends to a constant. For instance, in the limit $a\rightarrow 0$,
for $T=300$ [K] we have QFI for estimating $a$ as $\mathscr{F}_a\sim
6.16 \times 10^{-2} \;\mathrm{[1/km^2]}$, which gives a smallest
standard deviation of about $4$ [km]. Thus, we can conclude that the
photon number measurement on the complete basis of Fock states in the
detection mode $\hat b$ helps us to get better resolution than
heterodyne measurement. If $a$ becomes larger, we can see that the QFI
and CFI get close to each other at some point. To estimate  $a$, we
assumed that we know exactly the temperature of the source. 
Fig.~(\ref{figT:single}) shows that QFI for estimating the temperature
tends to zero in the limit $a\rightarrow 0 $. This 
demonstrates that one can not determine the temperature of infinitely small
sources with this method.

\begin{figure*}[t!]
	\centering
	\subfloat[][]{%
		\includegraphics[width=0.30\textwidth]{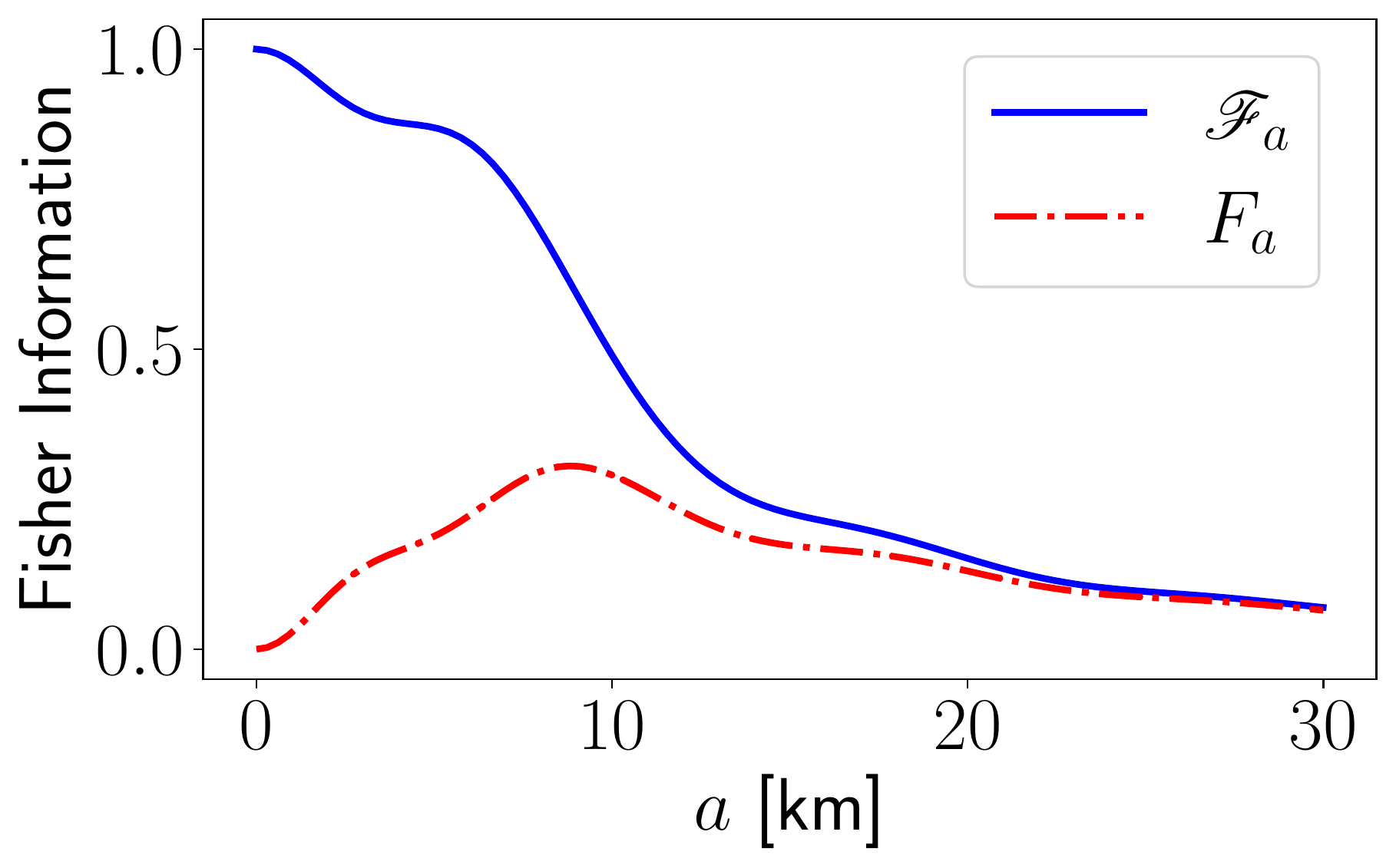}
		\label{figa:a}
	}
	\subfloat[][]{%
		\includegraphics[width=0.30\textwidth]{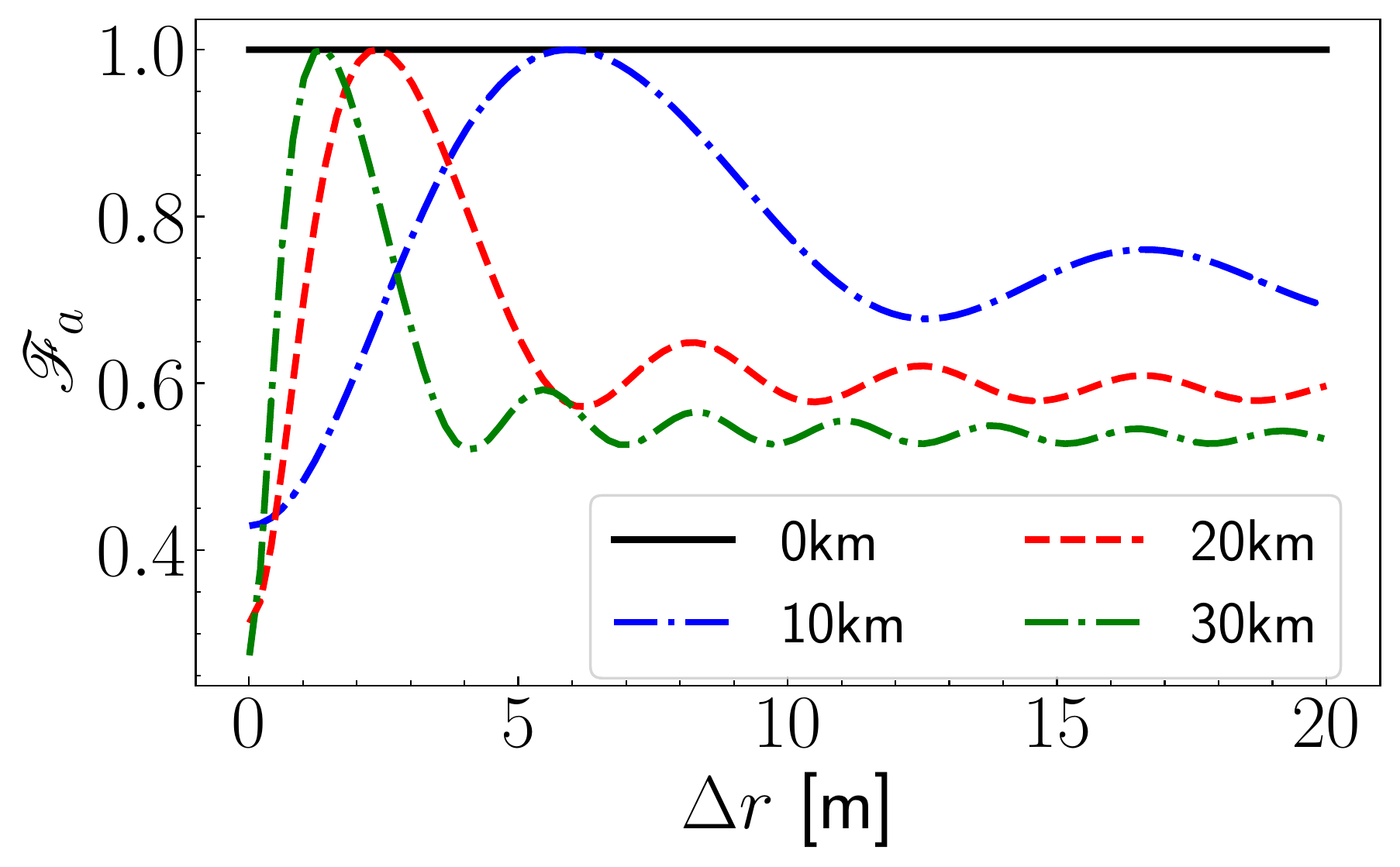}
		\label{figa2}
	}
	\subfloat[][]{%
		\includegraphics[width=0.295\textwidth]{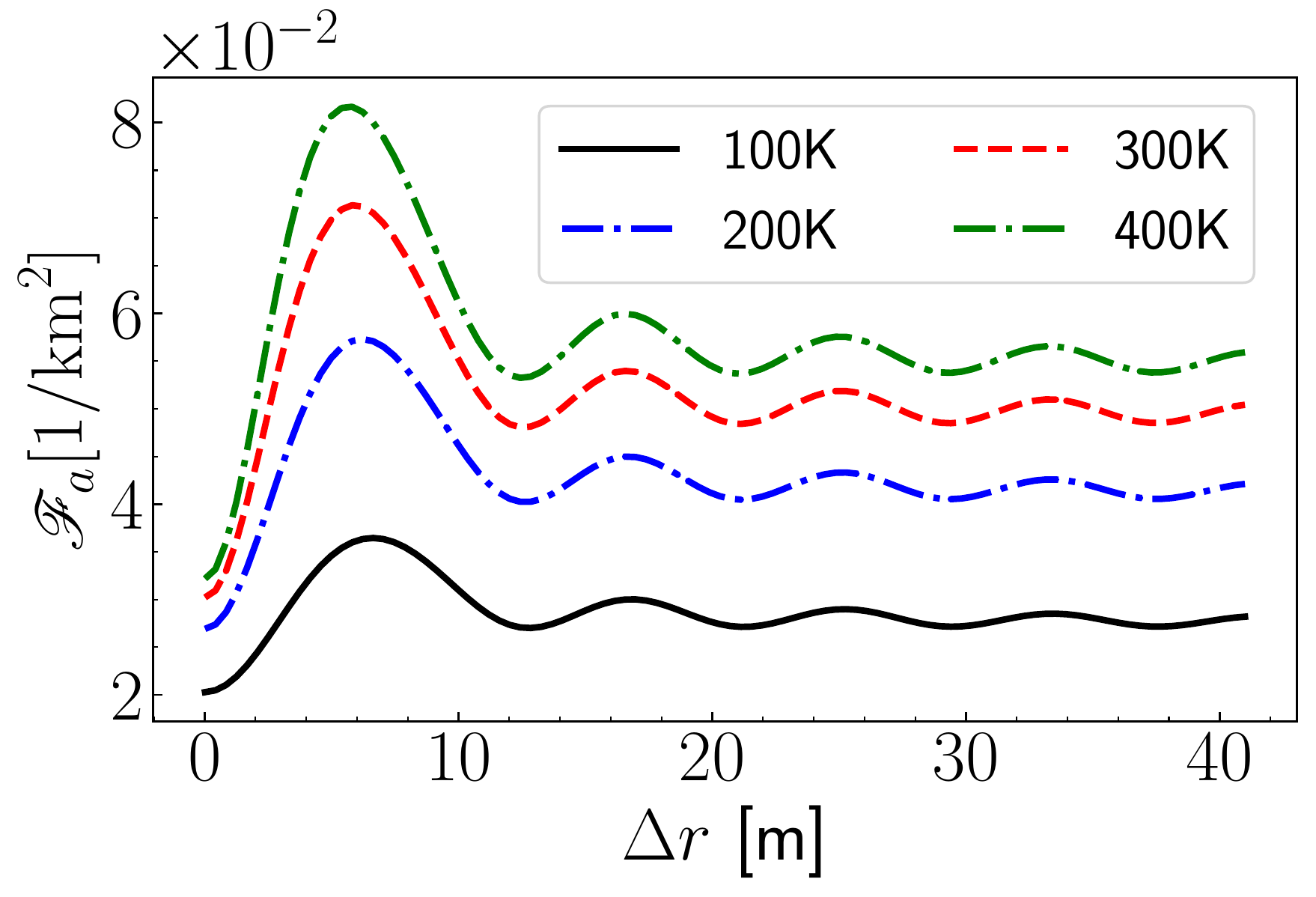}
		\label{figa:b}
	}
	\caption{a) The Fisher information concerning source size. The solid blue line shows the QFI for estimating the source size. The red dashed line shows the CFI from the heterodyne measurement. Both of the results are scaled with the Eq.~(\ref{Eqa}). In the limit $a\rightarrow 0$, for $T= 300$ [K], we have $\mathscr{F}_a \sim 0.117 \;[1/\mathrm{km}^2]$, which gives us a standard deviation around $2.92$ [km]. b) The QFI $\mathscr{F}_a$ for estimating the size of the circular disc as a function of $\Delta r$ (spatial separation of two receivers) for different source sizes $a$ with $T =300$ [K]. Each data scaled by the maximum value of the QFI for $a= (0,10,20,30)$ [km], which are $(\sim0.117,\sim0.070,\sim0.030,\sim0.015)[1/\mathrm{km}^2]$, respectively.  c) The QFI $\mathscr{F}_a$ for estimating the size of the circular disc as a function of separation of two receivers $\Delta r$, for different temperatures $T$.)}
	\label{figa}
\end{figure*}
Further, we can find the QFI for estimating the temperature as
\begin{equation}
	\begin{split}
		\mathscr{F}_T = \frac{\pi a^2 \kappa}{ R^2T+a^2\pi T^2 \kappa },
	\end{split}
\end{equation}
with a SLD given by
\begin{equation}
	\begin{split}
		\mathscr{L}_{T} = \frac{R^2}{TR^2 + a^2\pi T^2\kappa}\hat b^\dagger \hat b.
	\end{split}
\end{equation}
The CFI from heterodyne detection to estimate temperature becomes
\begin{equation}
	\begin{split}
		{F}_{T} = \frac{\pi^2 a^4 \kappa^2}{( R^2+a^2\pi T \kappa)^2 }.
	\end{split}
\end{equation}
In Fig.(\ref{figT:single}), we plot both QFI and CFI for heterodyne
detection for temperature estimation. Both have very close functional
behavior. They vanish for $a\rightarrow 0$ and they approach each
other when we have a large source size.

The off-diagonal matrix element of the QFI matrix for multiparameter estimation reads 
\begin{equation}
	\begin{split}
		\mathscr{F}_{aT} = \frac{2a\pi \kappa}{ R^2+a^2\pi T \kappa}.
	\end{split}
\end{equation}

{
  By sampling the same state 
  $\mathcal{N}$ times, the standard deviation of the estimator decreases proportional to  $1/\sqrt{\mathcal{N}}$. 
  The SMOS satellite moves at a constant speed $v\simeq 7$ [km/s] and takes the time $\tau = L/v$ to fly over a distance $L$. For each sample there is a lower bound for the detection time given by $t_D\sim 1/B $ (see appendix \ref{appendixAnew}). In practice, the effective detection time might be much larger, due to, e.g.~deadtimes of the detectors, slow electronics, etc.  In addition, zero temperature of the detector and modes $b_i$ is implicitly assumed in our calculations, but would require cooling down to temperatures much smaller than $\hbar \omega_0$.  If the actual detection time is $t_D^\text{eff}$, the sample size becomes $\mathcal{N}=\tau/t_D^\text{eff}$.  
  In this paper we intend to establish 
  the ultimate theoretical bounds and hence assume that the minimal detection time $t_D=1/B$ can be achieved, in which case the sample size becomes $\mathcal{N} \sim LB/v $. To estimate the source size one can assume that $L\sim a $, and the QCRB for estimating $a$ becomes $\delta a \geq 1/\sqrt{\mathcal{N}\mathcal{F}_a} $. Since $\mathcal{N}$ depends also on $a$ one can find the optimum bound in the sense of a minimal $\delta a$ at $a=R/\sqrt{\pi \kappa T}$. For $T= 300 $ [K], we find $a \sim 7.9$ [km] and $\delta a  \gtrsim  1.0$ [m]. 
  The bound for estimating $T$, assuming all other parameters known, can be written as $\delta T  \geq  1/\sqrt{\mathcal{N}\mathcal{F}_T} $. Using the same parameters as before and the same sample size, we have $\delta T \gtrsim 0.08$ [K]. Thus, increasing the sample size to the theoretically maximally possible value, the spatial resolution improves by a factor  of order 35,000 compared to the resolution of SMOS, and the radiometric resolution by factor of order 500. One can also increase the resolution by increasing the number of receivers, which we present in the following sections.}

\subsection{Two Mode Interferometer} In the previous section, we only
considered a single receiver with mode $\hat b$. It is obvious that we
may get additional information from the cross correlations of an $n$
mode interferometer. An analytical calculation of the QFI matrix for
$n$ mode interferometer generally becomes untractable for $n>2$ and
one has to rely on numerical calculation, see Section
\ref{sec3c}. Here, we consider the next simplest case of two receivers with modes $\hat{b}_1 $ and  $\hat{b}_2 $. Then we can write $\mathbf{b}$ as $ \mathbf{b}^{\top}=(\hat{b}_{1},\hat{b}_{1}^\dagger,\hat{b}_{2},\hat{b}_{2}^\dagger)$. The mean displacement is $\Gamma_{i} = 0$. The covariance matrix $\Sigma$ of the state $\rho_{\mathrm{int}}$ becomes

\begin{equation}
\begin{aligned}
\Sigma &=\left[\begin{array}{cccc}
0 & \chi & 0 & \xi \\
\chi & 0 & \xi^* & 0 \\
0 & \xi^* & 0 & \chi \\
\xi & 0 & \chi & 0
\end{array}\right],
\end{aligned}
\label{cov2}
\end{equation}
where $\chi =1/2+ \bar{n}  $ and $\xi = \braket{b_2^\dagger b_1} $. We give the general result for the QFI elements in Appendix~\ref{appendixA}. Further one can write the matrix $\mathbf{M}$ as
\begin{equation}
\mathbf{M}_i=\left[\begin{array}{cccc}
g^{1}_i & |g^{2}_i|e^{i\delta}  \\
|g^{2}_i|e^{-i\delta} & g^{1}_i
\end{array}\right],
\end{equation}
where $g^1_i, g^2_i$ are given in Appendix~\ref{appendixA} in terms of $\chi$ and $\xi$, and $\delta$ is the phase difference between two modes in the SLD. Using the eigenvectors of $\mathbf{M}$ we can write the unitary $\mathbf{V}$ as
\begin{equation}
\mathbf{V}=\frac{1}{\sqrt{2}}\left[\begin{array}{cccc}
1 & e^{i\delta}  \\
1 & -e^{i\delta}
\end{array}\right],
\label{eq:V}
\end{equation}
We see that $\mathbf{V}$ does not depend on the magnitude of the elements of the matrix $\mathbf{M}$ for a two-mode interferometer. The detection modes can be found as $\hat d_1 = (\hat b_1 +\hat b_2 e^{i\delta})/\sqrt{2}$ and $\hat d_1 = (\hat b_1 -\hat b_2 e^{i\delta})/\sqrt{2}$. The preprocessing to combine these two modes can be done by a phase delay on one mode and then combining these two modes by a beam splitter before any measurement. Then the POVM for the optimum measurement can be written as a set of projectors again in Fock basis as $\{\ket{m_1,m_2}\bra{m_1,m_2}\}_{\{m_1,m_2\}}$ which is the eigenbasis of  $\hat d^\dagger_i\hat d_i$, $\hat d^\dagger_i\hat d_i\ket{m_1,m_2}= m_i\ket{m_1,m_2} $. To compare this POVM with the classical approach, we consider heterodyne detection in Appendix~\ref{appendixB}.

\textit{\\\\Resolution of Uniform Circular Source:} Let us assume that on the source plane, we have a circular disk of radius $a$ with uniform temperature $T$ located at $\mathbf{r}=(x_0, y_0, R)$. Then the temperature distribution over the surface on the source plane can be written as 
\begin{equation}
    T_{\mathrm{eff}}(x,y) = T \mathrm{circ}({x-x_0,y-y_0}).
\end{equation}
We want to estimate again $a$ and $T$. The QFI for estimating the
source size is given by Eq.~(\ref{QFIa}) for a two-mode
interferometer. The expression is quite complicated. However, we can
analyze it numerically, or we can look at certain
limits. {Estimating the size of the circle
  $\mathscr{F}_{a}$ depends on $\Delta r $ (the separation of the two
  receivers). Physically we assumed this separation  to be greater than the central
  wavelength $\Delta r > \lambda $. Mathematically, one can take the limit
  $\Delta r \rightarrow 0 $, in which case the additional information
  from the phase difference between two receivers vanishes. In this
  limit, the QFI for estimating the source size becomes} 
\begin{equation}
\begin{split}
     \mathscr{F}_{a} \xrightarrow{\Delta r\rightarrow 0} \frac{8 \pi  \kappa T}{R^2+2\pi a^2\kappa T }.
\end{split}
\end{equation}
 If we have $2\pi a^2\kappa T \gg R^2 $, the QFI for estimating $a$ is ${\mathscr{F}_a}\sim 4/a^2$; for high temperatures or large $a$, the error of estimating the size of the source linearly increases with its size. Fig.~(\ref{figa2}) shows how the QFI changes when we decrease the source size. In the limit of $a\rightarrow 0 $, the QFI for estimating the source size becomes
\begin{equation}
\begin{split}
     \mathscr{F}_{a} \xrightarrow{a\rightarrow0} \frac{8 \pi  \kappa T}{R^2}.
     \label{Eqa}
\end{split}
\end{equation}
Comparing with the single receiver the QFI is doubled for two-mode
interferometers in the limit $a\rightarrow0$. We can still have
nonvanishing QFI for $a\rightarrow0$, as we can see from the black
line in Fig.~(\ref{figa2}), which is the limit as in
Eq.~(\ref{Eqa}). 
The black line ($\sim$0 [km] source size) is scaled with $\sim 0.117 \; [1/\mathrm{km}^2] $, which corresponds to a standard deviation of $\sim2.92$ [km] for the interferometer with two modes. We give the CFI for heterodyne detection in Eq.~(\ref{CFIa}). For small source size, we can ignore the higher-order terms in $a$, and we can simplify it as
\begin{equation}
	\begin{split}
		F_a \approx \frac{16 \pi^2 \kappa^2 T^2 a^2}{R^4}.
	\end{split}
\end{equation}
As we can see, for $a \rightarrow 0 $, the CFI for heterodyne
detection tends to zero.  Thus, the resolution of the source size with
heterodyne detection becomes arbitrarily bad in that limit (see
Fig.~(\ref{figa:a})). However, for large source sizes, we can see from
Fig.~(\ref{figa:a}) that CFI and QFI become equivalent. 
Therefore, constructing a POVM from the SLD can beat Rayleigh's resolution curse, even for estimating the source size. To construct the POVM for estimating the source size we give the elements of matrix $\mathbf{M}$ in Eqs.(\ref{ga1}-\ref{ga2}). The phase delay is found as {$\delta_a = x_0v_x + y_0v_y$,} with $v_i$ defined as $v_{x} = \Delta r \cos\varphi /(\lambda R)$, $v_{y} = \Delta r \sin\varphi /(\lambda R)$.
Thus, once we have the information of the location of the source
centroid, we can combine these two receivers modes by using a phase
delay to get the POVM that saturates the quantum Cramer Rao bound. {We
  plot the QFI as a function of $\Delta r$ in Fig.~(\ref{figa:b}) for
  different temperatures. We can see that when the effective
  temperature of the circular source increases, the QFI also
  increases. Moreover, when we increase $\Delta r $, the QFI for
  estimating $a$ increases up to a maximum around $\Delta r \sim 6$
  [m]. The reason for this is additional information coming from the
  phase differences in the two receivers. The QFI in Eq.~(\ref{Eqa})
  is doubled  compared to QFI for single receiver in
  Eq.~(\ref{Eqasingle}) in the limit $a\rightarrow 0$.} 
\begin{figure*}[t!]
	\centering
	\subfloat[][]{%
	\includegraphics[width=0.3\textwidth]{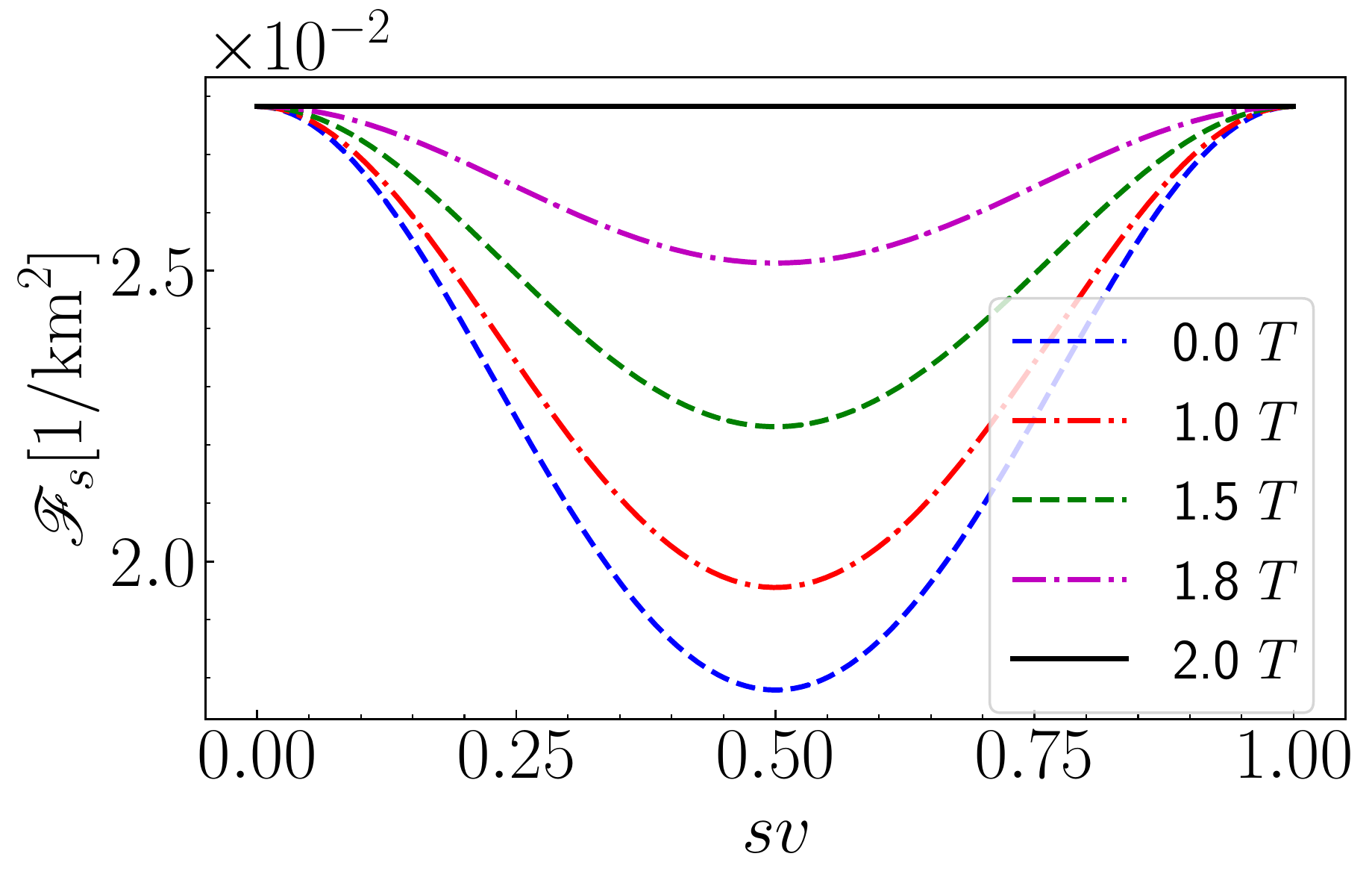}
	\label{figs:DT}
}
	\subfloat[][]{%
		\includegraphics[width=0.29\textwidth]{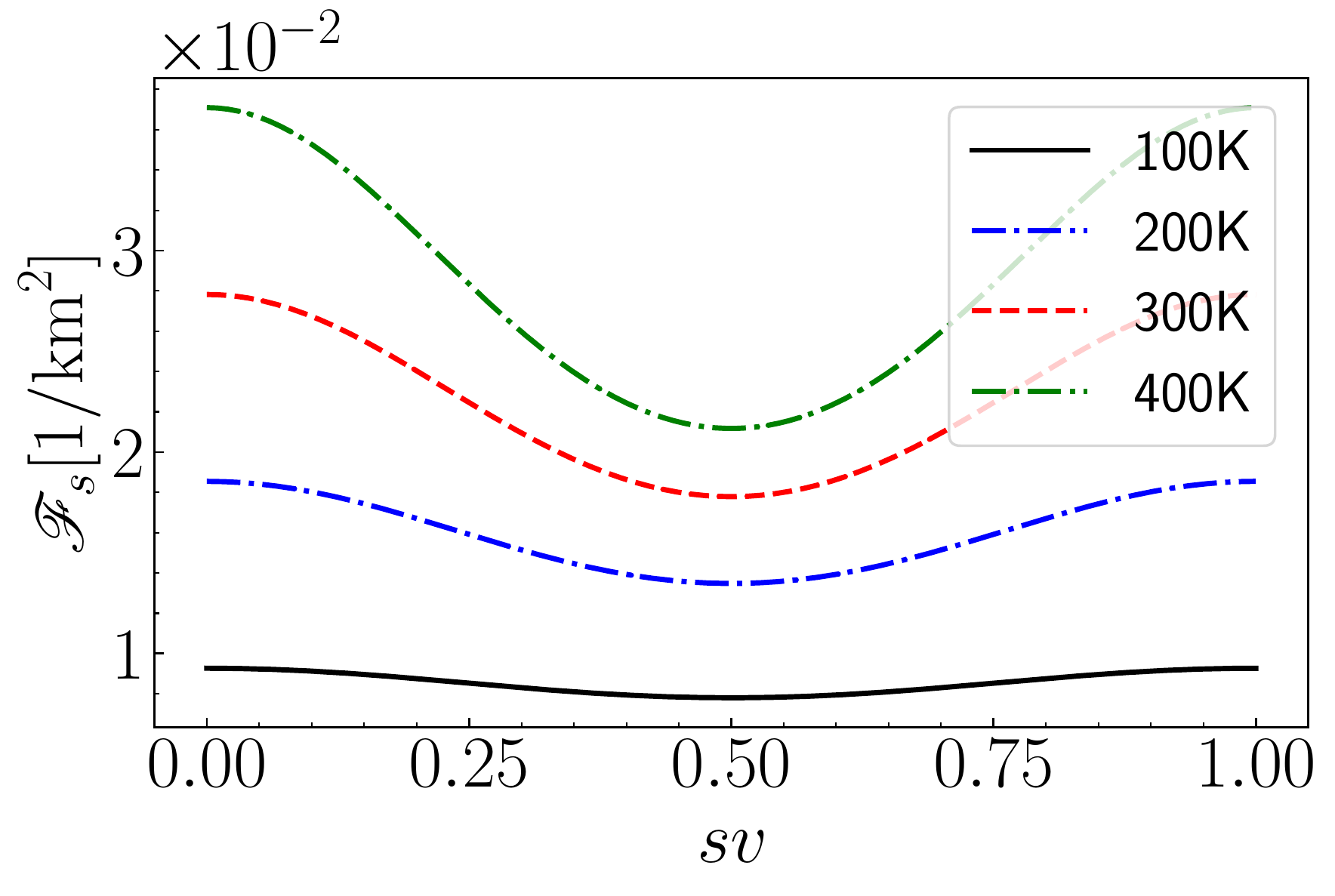}
		\label{figs:T}
	}
	\subfloat[][]{%
		\includegraphics[width=0.3\textwidth]{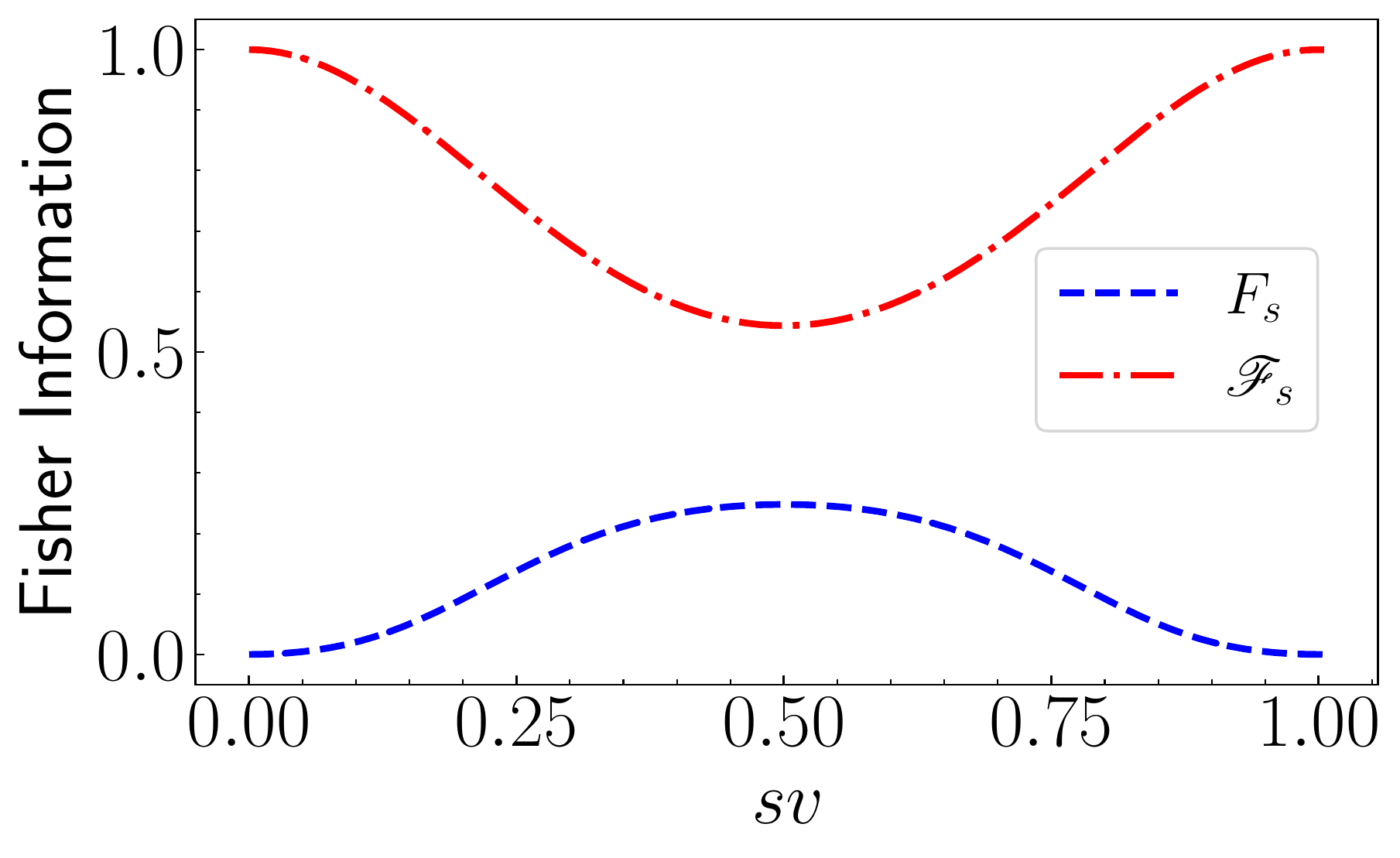}
		\label{figs:CFI}
	}
	\\
	\subfloat[][]{%
	\includegraphics[width=0.3\textwidth]{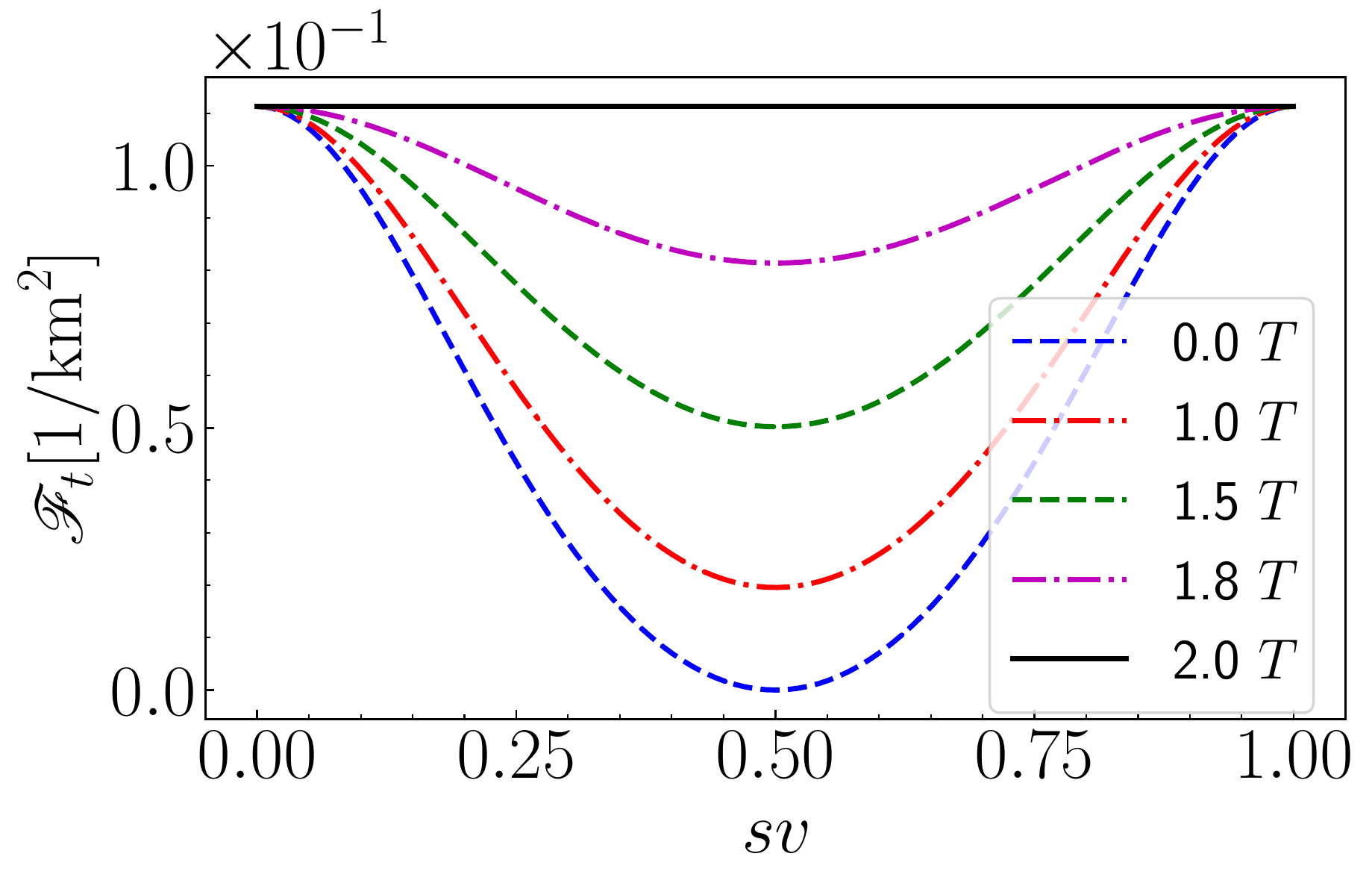}
	\label{figt:DT}
}
	\subfloat[][]{%
	\includegraphics[width=0.3\textwidth]{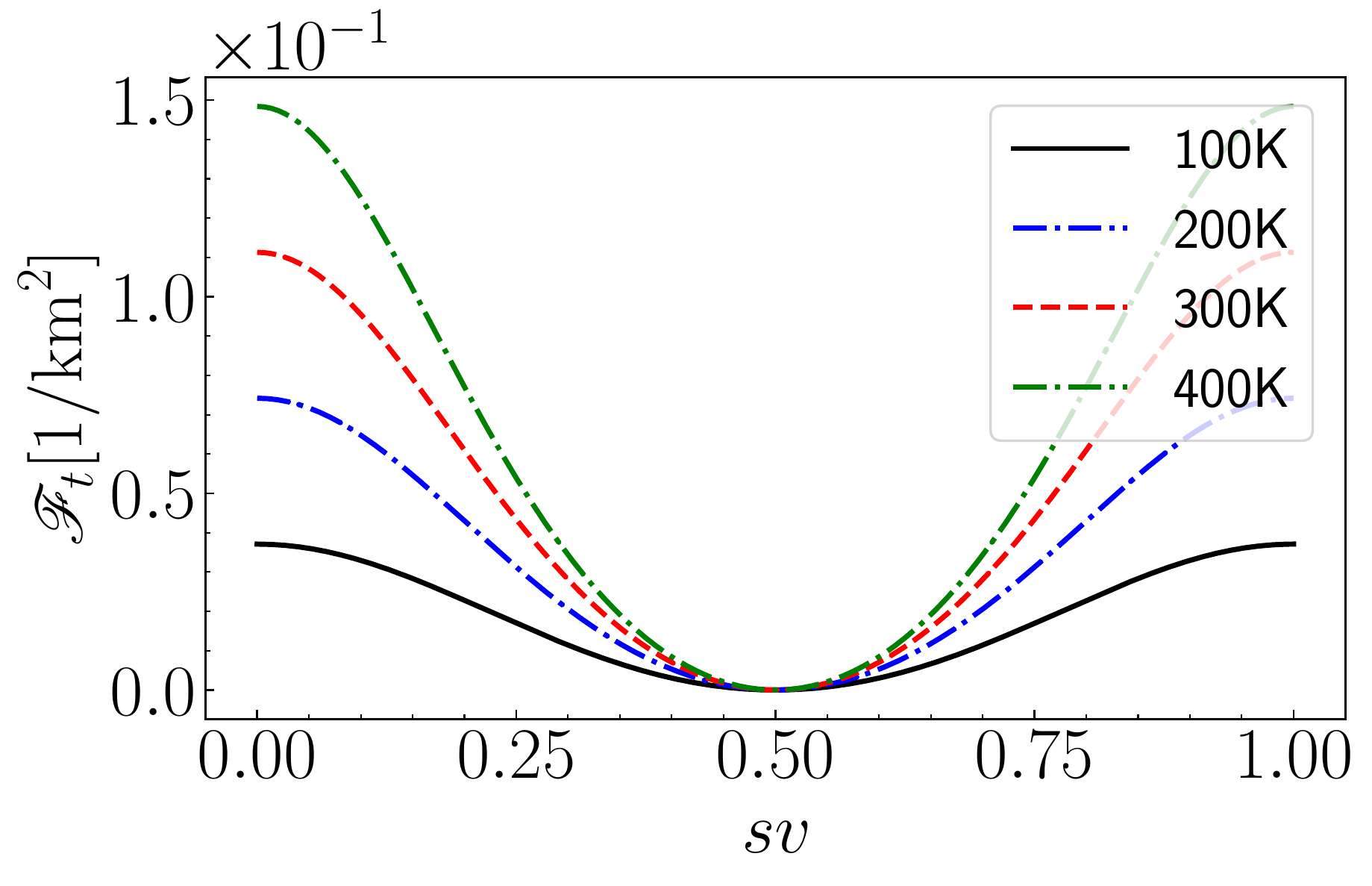}
	\label{figt:T}
	}
	\subfloat[][]{%
	\includegraphics[width=0.3\textwidth]{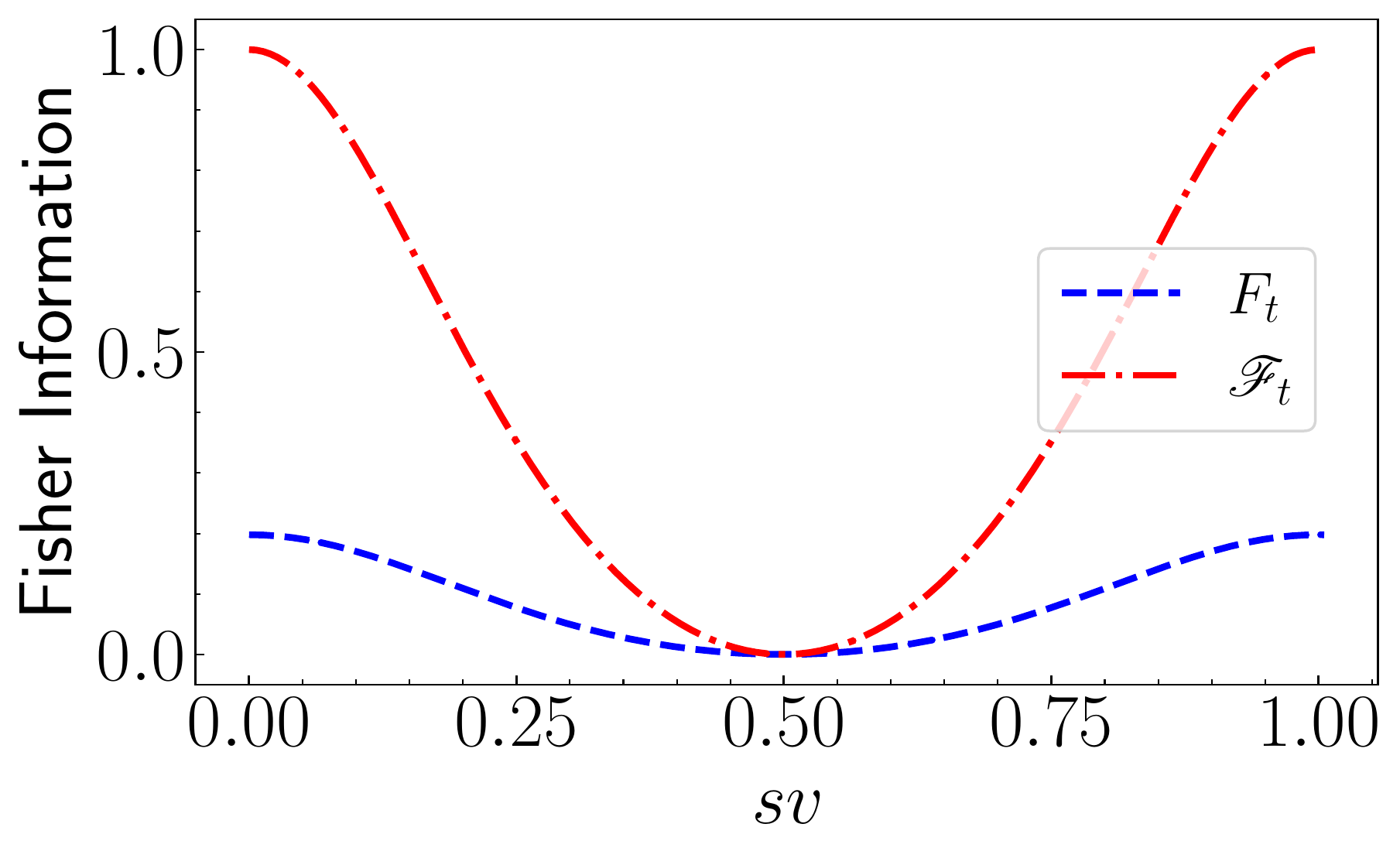}
	\label{figt:CFI}
	}
	\caption{a) The QFI $\mathscr{F}_s$ with respect to $sv $ for various temperature difference $\Delta T$ for $T = 300$ [K]. b) The QFI of estimating the separation of two-point sources as a function of $sv $ for different average temperatures. c) The QFI $\mathscr{F}_s$ (red dot-dashed) and CFI for heterodyne detection (Blue dashed) for estimating the separation of two-point sources as a function of $sv$. Both the QFI and the CFI scaled by $4\pi^2 v^2 \eta \kappa T$. d) The QFI $\mathscr{F}_t$ respect to $sv $ for various temperature difference $\Delta T$  for $T = 300$ [K]. e) The QFI $\mathscr{F}_t$ respect to $sv $ for different average temperature $T$. For (a), (b), (d), and (e), the separation $\Delta r$ of the two receivers is fixed by 4 [m] and $\eta \sim 10^{-4}$. f) The QFI $\mathscr{F}_t$ (red dot-dashed) and CFI for heterodyne detection ${F}_t$ (Blue dashed) for estimating the separation of two-point source as a function of $sv$. Both the QFI and the CFI scaled by $32\pi^2 v^2 \eta \kappa T$.}
\end{figure*}

In the limit $\Delta r \rightarrow 0 $, the QFI for estimating $T$ becomes
\begin{equation}
    \begin{split}
        \mathscr{F}_{T} \rightarrow \frac{2 \pi  a^2 \kappa  }{T \left(2 \pi  a^2 \kappa T+R^2\right)}.
    \end{split}
\end{equation}
Since we assume we are in a microwave regime $k_BT \gg \hbar \omega_0$, we can not take the limit $T\rightarrow 0 $. Instead, we can verify that the QFI for estimating the temperature depends on the source size for a finite temperature. Now, for $T=$300 [K], and $30$ [km] source size we have the QFI around $2\times 10^{-5} $ [$1/\mathrm{K}^2$] which gives a very high standard deviation around 221 [K]. We show in the next section that the QFI also increases if we increase the number of spatial modes. For instance, for 20 receivers, we have QFI around $1.5\times 10^{-4} $ [$1/\mathrm{K}^2$], and the standard deviation is 79 [K] for a single measurement.

In the limit $\Delta r \rightarrow 0 $, the CFI from heterodyne detection becomes
\begin{equation}
	\begin{split}
		F_T \rightarrow \frac{4 \pi ^2 a^4 \kappa ^2 \left(\pi  a^2 \kappa  T+R^2\right) \left(3 \pi  a^2 \kappa  T+R^2\right)}{\left(2 \pi  a^2 \kappa  T+R^2\right)^4}.
	\end{split}
\end{equation}
To compare with the QFI we assume the brightness temperature $T=$ 300 [K], and source size $a=$ 30 [km]. This gives a CFI around $8\times 10^{-6} $ [$1/\mathrm{K}^2$] which give us a standard deviation around 350 [K]. Compared to the QFI information, the CFI is around 2.5 times smaller. Therefore, combining the spatial modes (receivers) and measuring photon number in the Fock basis of $\hat d_1, \hat d_2$, as expected, is more advantageous even for estimating the temperature.

So far, we only gave the diagonal elements of the QFI matrix, relevant
for estimating each parameter individually, assuming all other parameters are known. The single independent off-diagonal element of the QFI matrix regarding $a$ and $T$ is given in Eq.~(\ref{QFIaT}). In the limit $\Delta r\rightarrow 0$ it simplifies to 
\begin{equation}
	\mathscr{F}_{aT} = \frac{4\pi  a \kappa }{2 \pi  a^2 \kappa  T+R^2}.
\end{equation}
Then one can construct the QFI matrix to find the quantum Cram\'er-Rao bound for multiparameter estimation. Further, we can estimate the source location considering the two parameters $x_0,y_0$. The QFI matrix elements for estimating the source locations can be written as 
\begin{equation}
	\begin{split}
		\mathscr{F}_{i_0j_0} = \frac{8 \pi ^2 \Delta r^2  \kappa  T J_1{}^2 v_i v_j}{\pi  {\Delta r}^2 \left(\pi  a^2 \kappa  T+R^2\right)-\kappa  \lambda ^2 R^2 T J_1{}^2},
	\end{split}
\end{equation}
where $i,j \in \{x,y\}$. The QFI for estimating the source location
depends on source size and source temperature. Since the elements
$\mathscr{F}_{i_0a}$ and $\mathscr{F}_{i_0T} $ of the QFI matrix  are zero, source size and location can be estimated simultaneously. And the necessary phase delay for POVM can be found as $\delta _{i_0} =\delta+\pi/2 $ from Eq.~(\ref{g2ij}).

\begin{figure*}[t!]
	\centering
	\subfloat[][]{%
	\includegraphics[width=0.375\textwidth]{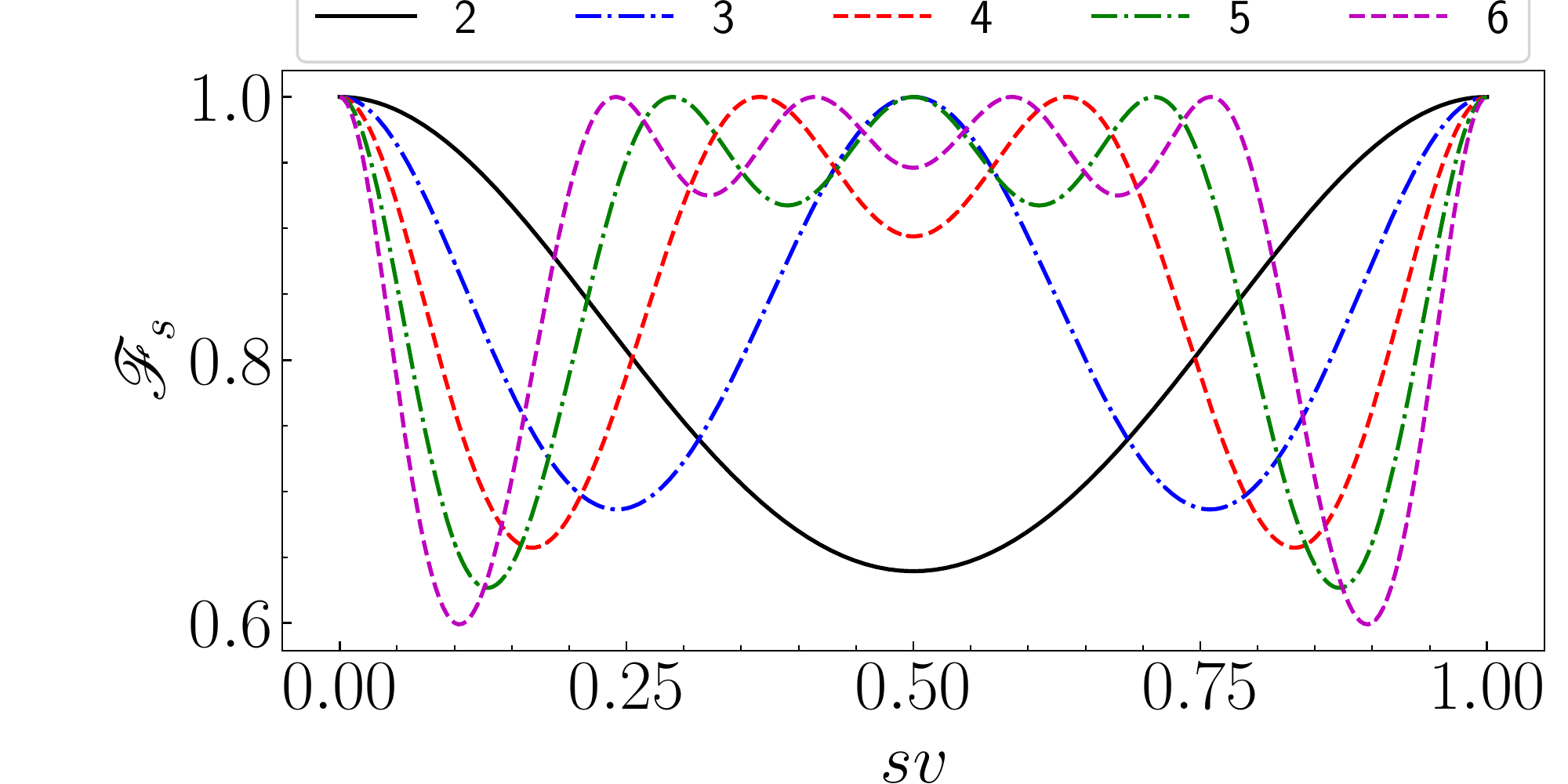}
	\label{n:s6}
	}
	\subfloat[][]{%
		\includegraphics[width=0.28\textwidth]{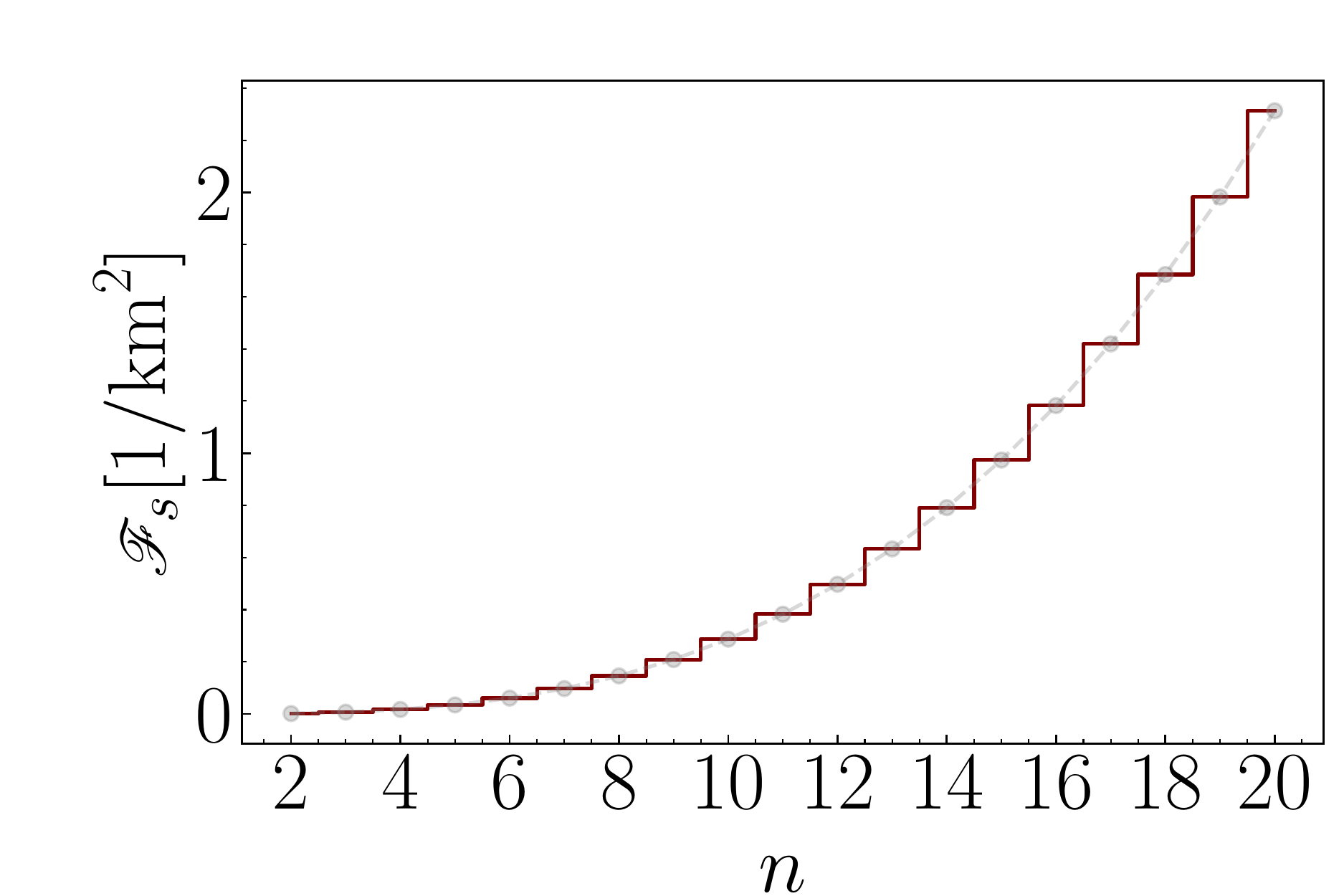}
		\label{n:s20}
	}
	\subfloat[][]{%
		\includegraphics[width=0.28\textwidth]{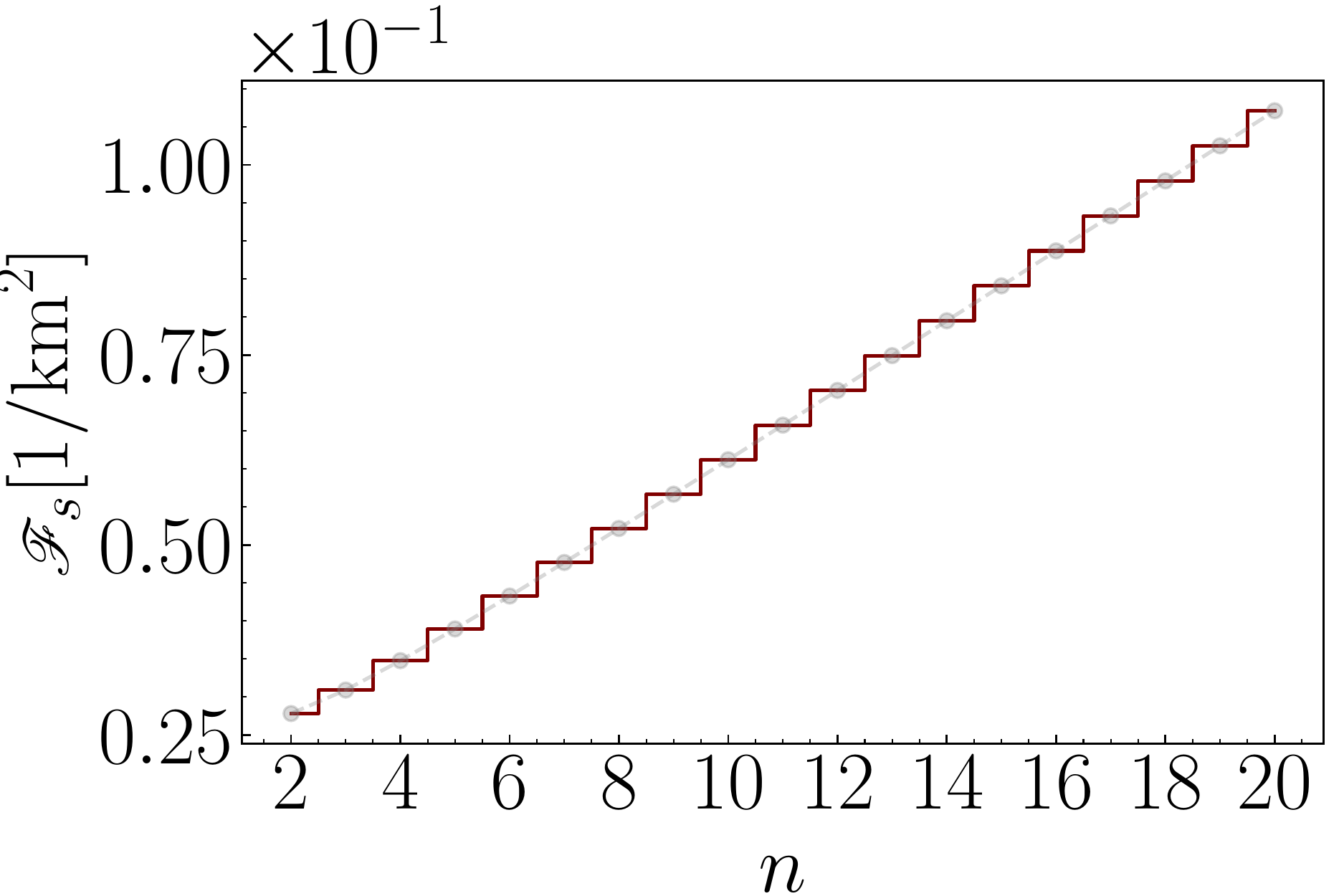}
		\label{n:s20fixed}
	}

	\subfloat[][]{%
	\includegraphics[width=0.375\textwidth]{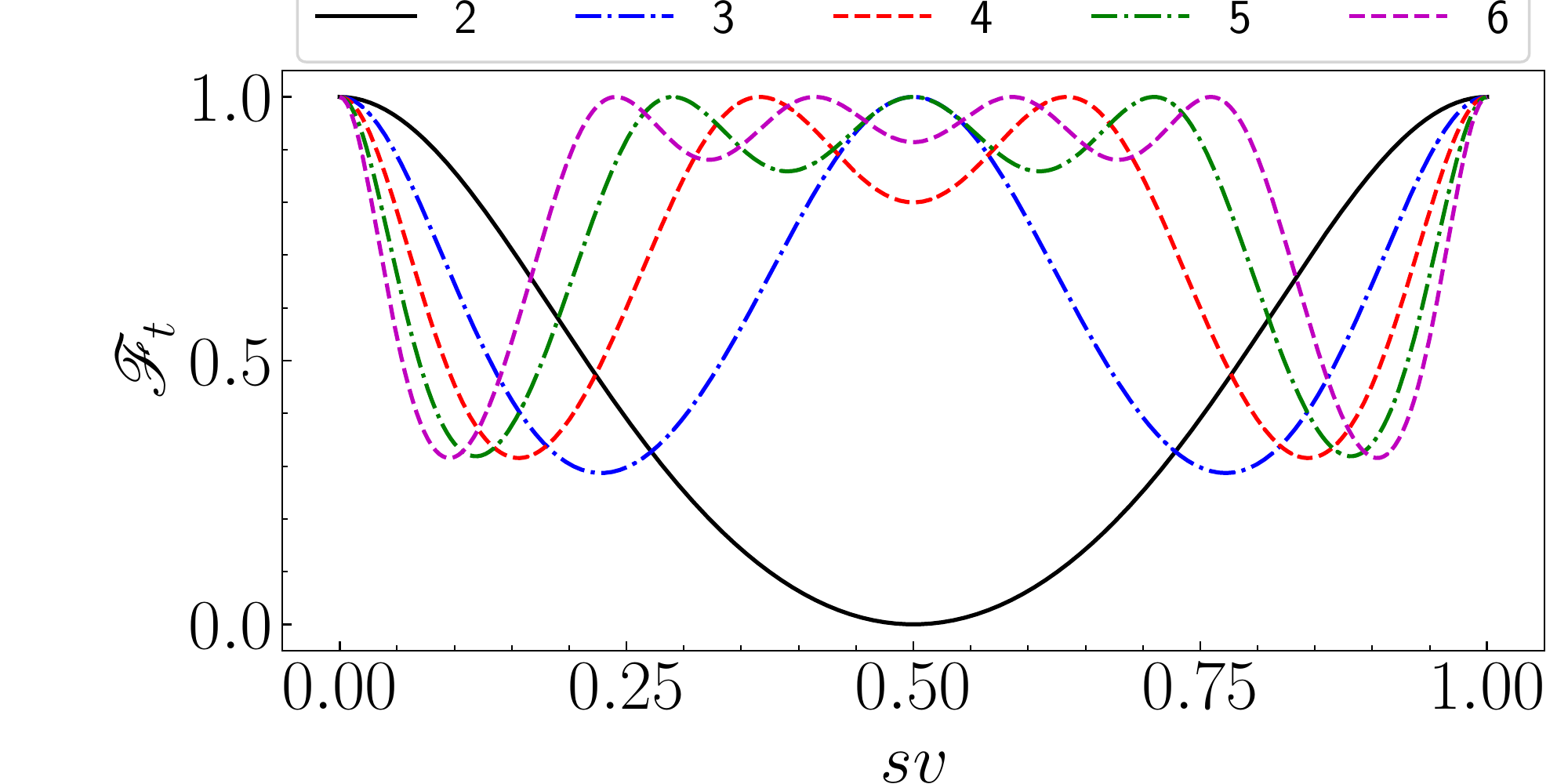}
	\label{n:t6}
    }
	\subfloat[][]{%
		\includegraphics[width=0.28\textwidth]{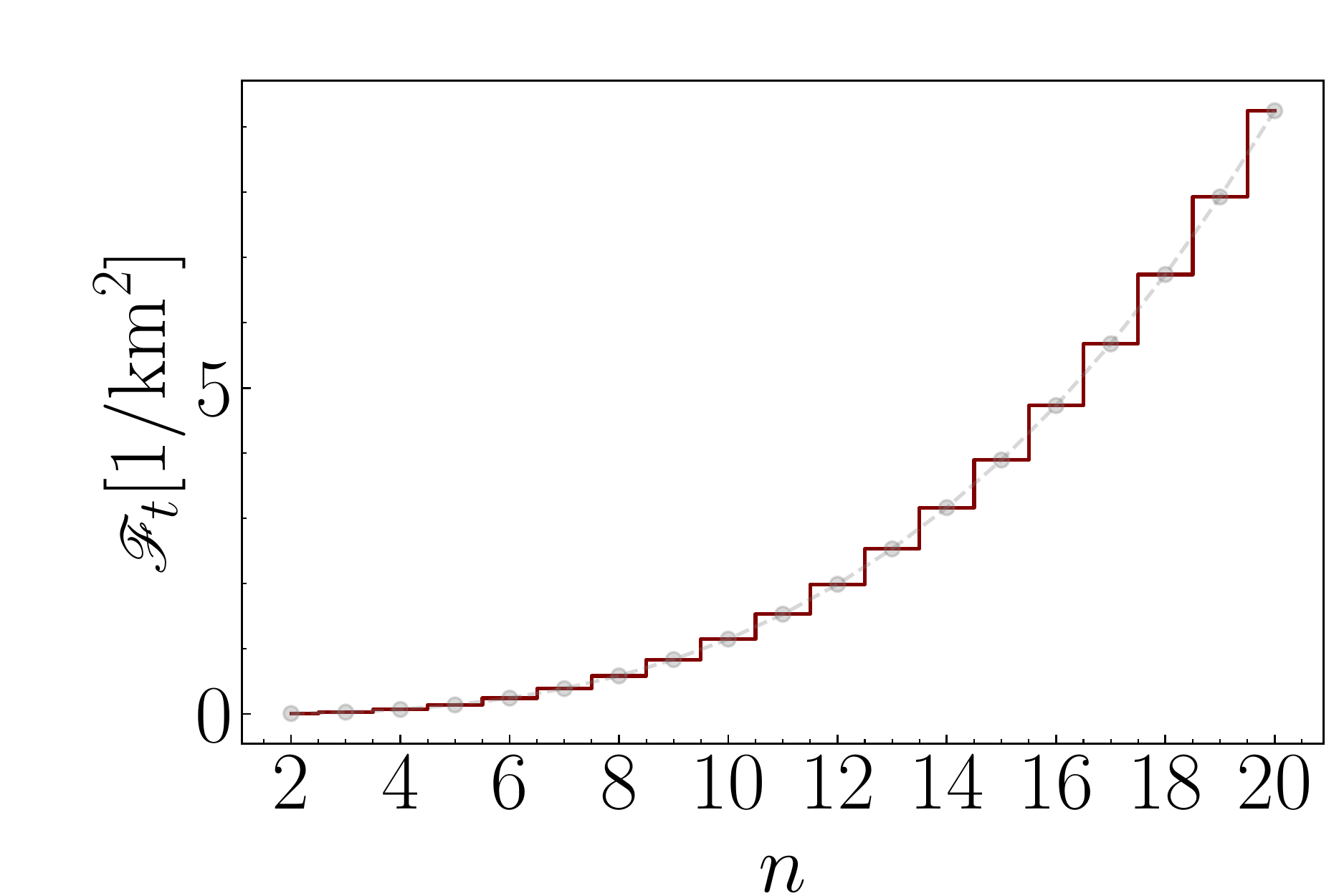}
		\label{n:t20}
	}
	\subfloat[][]{%
		\includegraphics[width=0.28\textwidth]{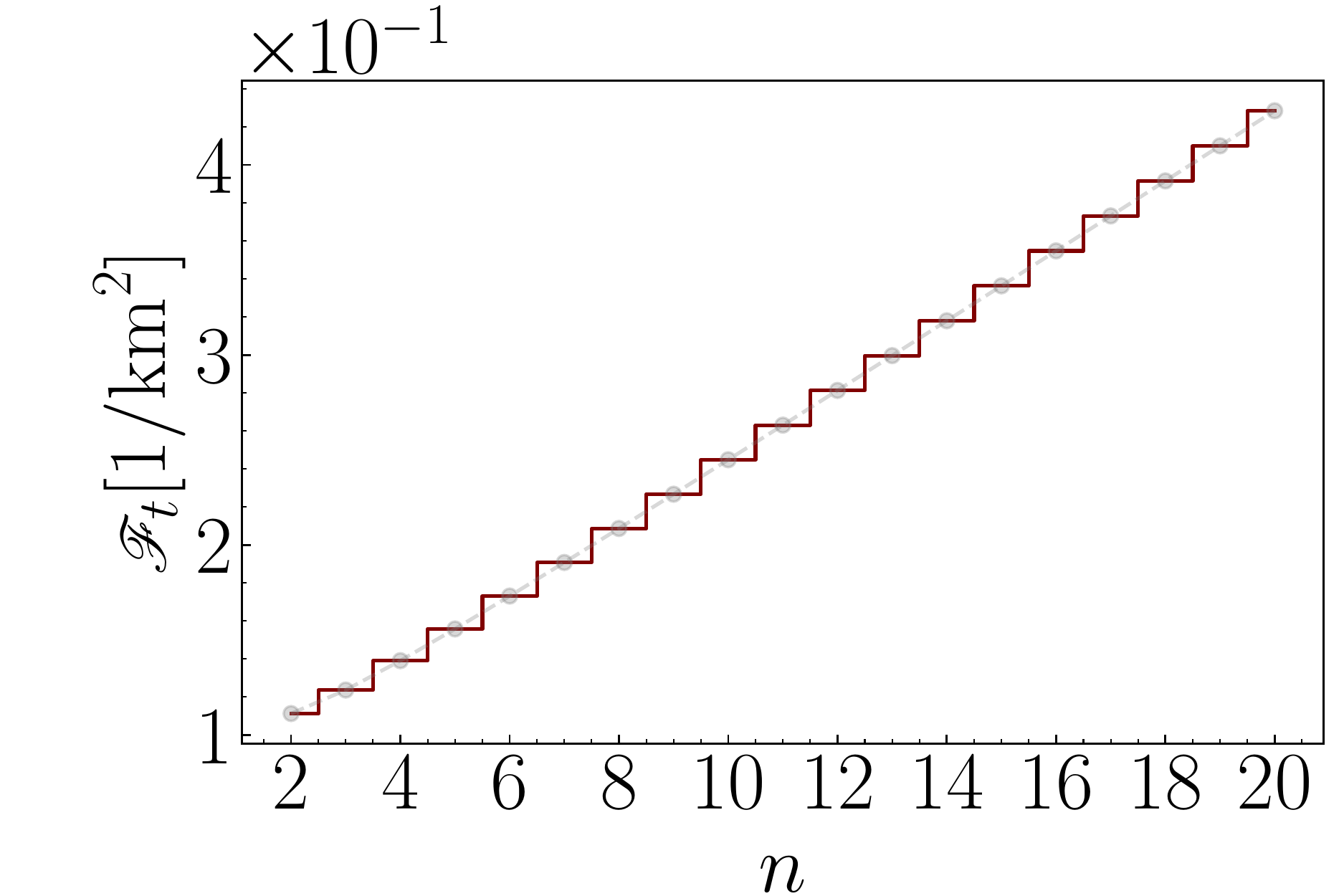}
		\label{n:t20fixed}
	}
	\caption{a) The QFI $\mathscr{F}_s$ as a function of $sv $ for $(2,3,4,5,6)$ mode interferometers and each curve is scaled by its maximum value which are $(\sim 1.7\times 10^{-3},\sim 6.9\times 10^{-3},\sim 17.3\times 10^{-3},\sim 34.7\times 10^{-3},\sim 60.8\times 10^{-3})[1/\mathrm{km}^2]$, respectively. b) The QFI $\mathscr{F}_s$ with respect to a number of interferometer modes $n$. d) The QFI $\mathscr{F}_t$ as a function of $sv $ for $(2,3,4,5,6)$ mode interferometers. Each curve is scaled by its maximum value which are $(\sim 0.67\times 10^{-2},\sim 2.78\times 10^{-2},\sim 6.96\times 10^{-2},\sim 13.9\times 10^{-2},\sim 24.3\times 10^{-2})[1/\mathrm{km}^2]$, respectively. e) The QFI $\mathscr{F}_t$ with respect to a number of interferometer modes $n$. (For all, the separation of two nearest receiver $\Delta r$ is 1m, and $\eta \sim 10^{-4}$. The maximum baseline is $ \Delta r_{\mathrm{max}}=(n-1)\Delta r$.) c) The QFI $\mathscr{F}_s$ with respect to a number of interferometer modes $n$. f) The QFI $\mathscr{F}_t$ with respect to a number of interferometer modes $n$. For both (c) and (f), the maximum baseline is fixed by $\Delta r_{\mathrm{max}}=4$ [m], in this case, separation of two nearest receiver is $\Delta r = \Delta r_{\mathrm{max}}/(n-1)$, and $\eta \sim 10^{-4}$. }

\end{figure*}

\textit{Spatial Resolution of Two Point Sources:} Recently, the spatial resolution of two equally bright strong point sources was studied in \cite{wang_superresolution_2021} by considering the sources aligned parallel to the two-mode interferometer.

In this section, we consider a similar model with two circular disc sources on the surface of the source plane at locations ($\mathbf{r}_1= (x_1,y_1,R)$ and $\mathbf{r_2} = (x_2,y_2,R)$) but with different effective temperatures $T_1$ and $T_2$, and same sizes $a$. We assume that in the far field $\{x_i,y_i\}\ll R$ and $a\ll R$. We analyze two cases; when the sources are aligned or not aligned with the two receivers. For two circular sources with equal size, the temperature distribution over the surface can be written as
\begin{equation}
    T_{\mathrm{eff}}(x,y) = \sum_{i=\{1,2\}} T_i \mathrm{circ}({x-x_i,y-y_i}).
	\label{twotemp}
\end{equation}

Then we can define the four parameters that we want to estimate as: source separation ($s_x = x_1 - x_2$), ($s_y = y_1 - y_2$) and centroid of the two sources ($ t_x = (x_1 + x_2)/2 $), ($ t_y = (y_1 + y_2)/2 $). In Appendix~\ref{appendixD}, we express the QFI matrix elements for all four parameters. Since these equations are quite complicated, we check the important limits. Since we want to resolve the two-point sources even for very small separation, we check the limit, $s_x,s_y \rightarrow 0$. Then we have QFI matrix elements for estimating the source separation as $\mathscr{F}_{s_i} \rightarrow{4\pi^2 v_i ^2 \eta \kappa T }$ and $ \mathscr{F}_{s_xs_y} \rightarrow{4\pi^2 v_x v_y \eta \kappa T }$, if $T_1 = T_2 = T$.

If we assume two sources aligned parallel to the two-mode interferometer ($y_1=0 , y_2=0$ and $\varphi = 0\rightarrow v=v_x=\Delta r /(R\lambda) $ and $s_x\rightarrow s,\quad t_x\rightarrow t $) we can simplify our problem to a single dimension. We show the dependence of the QFI matrix elements on average temperature ($T= (T_1 +T_2)/2$) and temperature difference $\Delta T =T_1- T_2$ assuming $T_1 \geq T_2$. In Fig.~(\ref{figs:T}), we plot $\mathscr{F}_s$ with respect to $sv $ for different average temperatures.
As expected, when the temperature increases, the QFI for estimating the separation also increases. For $T=300$ [K] and $\Delta r = 4$ [m], we have a QFI around $0.027 \;[1/\mathrm{km}^2]$ which corresponds to a standard deviation of 6 [km] for only two receivers for the separation estimation. In Fig.~(\ref{figs:DT}), we see that, as we increase the temperature difference between the two-point sources, the QFI becomes less oscillatory and at $\Delta T\rightarrow 2T$, the oscillatory behavior disappears. In the limit $\Delta T \rightarrow 2T$, or $s\rightarrow 0 $ the QFI for estimating $s$ becomes
\begin{equation}
\begin{split}
   \mathscr{F}_s \rightarrow{4\pi^2 v ^2 \eta  \kappa  T }\,,
\end{split}
\end{equation}
 which is the limit given by the solid black line in Fig.~(\ref{figs:DT}). We calculated the CFI from heterodyne detection to estimate the source separation in Eq.~(\ref{CFIs}). If the size of the sources is very small and in the limit $\eta  \kappa  T \ll 1$ the CFI for estimating the source separation simplifies to
 \begin{equation}
	 \begin{split}
		 F_{s_i} \xrightarrow{\eta  \kappa  T \ll 1} 8 \pi ^2 \eta ^2 \kappa ^2 T^2 v_i^2 \sin ^2(\pi  (s_xv_x +s_yv_y)).
	 \end{split}
 \end{equation}
 When the source separation goes to zero ($s_x,s_y\rightarrow 0$), $F_{s_i}$ tends to zero. We compare the QFI with CFI in Eq.~(\ref{CFIs}) from heterodyne detection in Fig.~(\ref{figs:CFI}). As we can see, the CFI goes to zero for small source separation. Therefore, we can conclude that Rayleigh's curse limits heterodyne detection. The POVM from the SLD eliminates that limitation. We give the elements of the matrix $\mathbf{M}_s$, $g^1_s$ and $g^2_s$ in Appendix \ref{appendixD}. The phase difference for combining two spatial modes of the interferometer can be found as $\delta_s = 2\pi (t_x v_x+t_yv_y)-\pi$. Assuming the alignment of the spatial mode separation parallel to source separation, it becomes $\delta_s = 2\pi t v-\pi$.

 {
 The QFI matrix elements for estimating the centroid is given in Eq.~(\ref{QFIt}). We assume that the two sources aligned again parallel to two spatial modes of the
 interferometer ($y_1=0 , y_2=0$ and $\varphi = 0\rightarrow v=v_x=\Delta r /(R\lambda) $ and $s_x\rightarrow s,\quad t_x\rightarrow t $) and $\mathscr{F}_{t_{x},t_{x}}\rightarrow \mathscr{F}_{t}$.
In Fig.~(\ref{figt:T}), we see that the $\mathscr{F}_t$ increases when
we increase the temperature. For $T =300$ [K] and $\Delta r = 4 $ [m],
we have a QFI $\mathscr{F}_t\sim 0.11\; [1/\mathrm{km}^2]$ which corresponds to a standard deviation of 3 [km] for estimating the centroid. When $sv \sim 0.5 $,  $\mathscr{F}_t $ goes to zero for equally bright sources. In Fig.~(\ref{figt:DT}) we see that it is not zero for $sv \sim 0.5 $, if $\Delta T \neq 0 $, and the oscillation of $\mathscr{F}_t$ decreases when we increase the temperature difference. In the limit $s\rightarrow 0$, $\mathscr{F}_t$ simplifies to
\begin{equation}
	\begin{split}
	   \mathscr{F}_t \rightarrow{32 \pi^2 v ^2 \eta  \kappa  T }
	\end{split}
\end{equation}
for $\Delta T = 0 $.} The CFI for heterodyne detection is given in Eq.~(\ref{CFIt}). For small sources we consider again the limit $\eta  \kappa  T \ll 1$, and we ignore the higher order terms of $\eta  \kappa  T $. Then we have
\begin{equation}
	\begin{split}
		F_{t_i} \xrightarrow{\eta  \kappa  T \ll 1} 32 \pi ^2 \eta ^2 \kappa ^2 T^2 {v_i}^2 \cos ^2(\pi  ({s_x} {v_x}+{s_x} {v_y})).
	\end{split}
\end{equation}
 If the source separation goes to zero ($s_x,s_y\rightarrow 0$), we still have a finite $F_{t_i}$, unlike the CFI for source separation. In Fig.~(\ref{figt:CFI}), we compare $\mathscr{F}_t$ with $F_{t_i}$. When the source separation goes to zero, both Fisher information goes to a constant, and both go to zero at $sv\rightarrow 0.5$. However, the QFI is five times larger than the CFI from heterodyne detection. Again the phase difference for the POVM from the SLD can be found as $\delta_t = 2\pi tv +\pi/2$.

\subsection{1D $n$-mode interferometer arrays}
\label{sec3c}
The previous section considered a two-mode interferometer for analytical calculations and compared the QFI with its POVM and CFI for heterodyne detection. To compare our results with SMOS, we extend the two-mode interferometer to a 1D array of $n$ single-mode receivers. We investigate numerically how the QFI changes when increasing the number $n$ of interferometer modes. We assume the receiver array aligned with the $x$ axis on the detection plane and denote the maximum baseline separation of the two most distant receivers by $\Delta r_{\mathrm{max}}$.

\textit{Resolution of Two Point Sources for n mode interferometer array:}
We assume that both sources have the same sizes and temperatures ($\Delta T = 0 $ and $T_1 = T_2 = T$) and that they are parallel to the receiver array. In Fig.~(\ref{n:s6}), we see that when we increase the number of receivers, the behavior of $\mathcal{F}_s$ changes. It is still oscillatory as a function of $ sv $ with a period of $2\pi$. However, for each oscillation, we have $n-2$ additional maxima. Moreover, in Fig.~(\ref{n:s20}), we see that $\mathcal{F}_s$ increases gradually when we increase the number of receivers and the maximum baseline increases as $\Delta r_{\mathrm{max}} = (n-1)\Delta r$. For $\Delta r = 1 $ [m] and $T=300$ [K], the standard deviation for estimating the source separation is $\sim 23$ [km] for the two-mode interferometer. For the 20 mode interferometer, we find a standard deviation of around $0.65$ [km]. Further, if we keep the baseline fixed as $4$ [m], the QFI increases linearly with the number of receivers, as we can see in Fig.~(\ref{n:s20fixed}). In this case, for a two-mode interferometer, we have a standard deviation of around $6 $ [km], and for a 20-mode interferometer, we have $3 $ [km].

We also checked the centroid estimation for the $n$ mode interferometer. It leads to similar results as for source separation. From Fig.~(\ref{n:t6}) we see that for $sv = 0.5$ the centroid uncertainty for the two-mode receiver diverges ($\mathscr{F}_t \sim 0 $ at $sv \sim 0.5 $). This is no longer the case for the array of $n$ receivers. In Fig.~(\ref{n:t20}), we see that $\mathscr{F}_t$ also increases with the number of modes. For the two modes, the standard deviation for estimating the centroid was $\sim 12$ [km]. For the 20 modes, we have a standard deviation of around 0.32 [km] considering $\Delta r = 1 $ [m], and $\Delta r _\mathrm{max} = (n-1)\Delta r$ for average temperature $T= 300$ [K]. If we keep the baseline fixed, as we can see from the Fig.~(\ref{n:t20fixed}), $\mathscr{F}_t $ increases linearly by $n$. By fixing the $\Delta r _\mathrm{max} = 4$ [m], we have a standard deviation of $\sim 3$ [km] for the two mode interferometer; for 20 modes we have a standard deviation of $\sim 1.5$ [km]. Thus, instead of sampling the state in time, we can increase the number of receivers to increase the QFI, and both methods can be combined as well.

\begin{figure}[h!]
	\centering
	\includegraphics[width=0.9\linewidth]{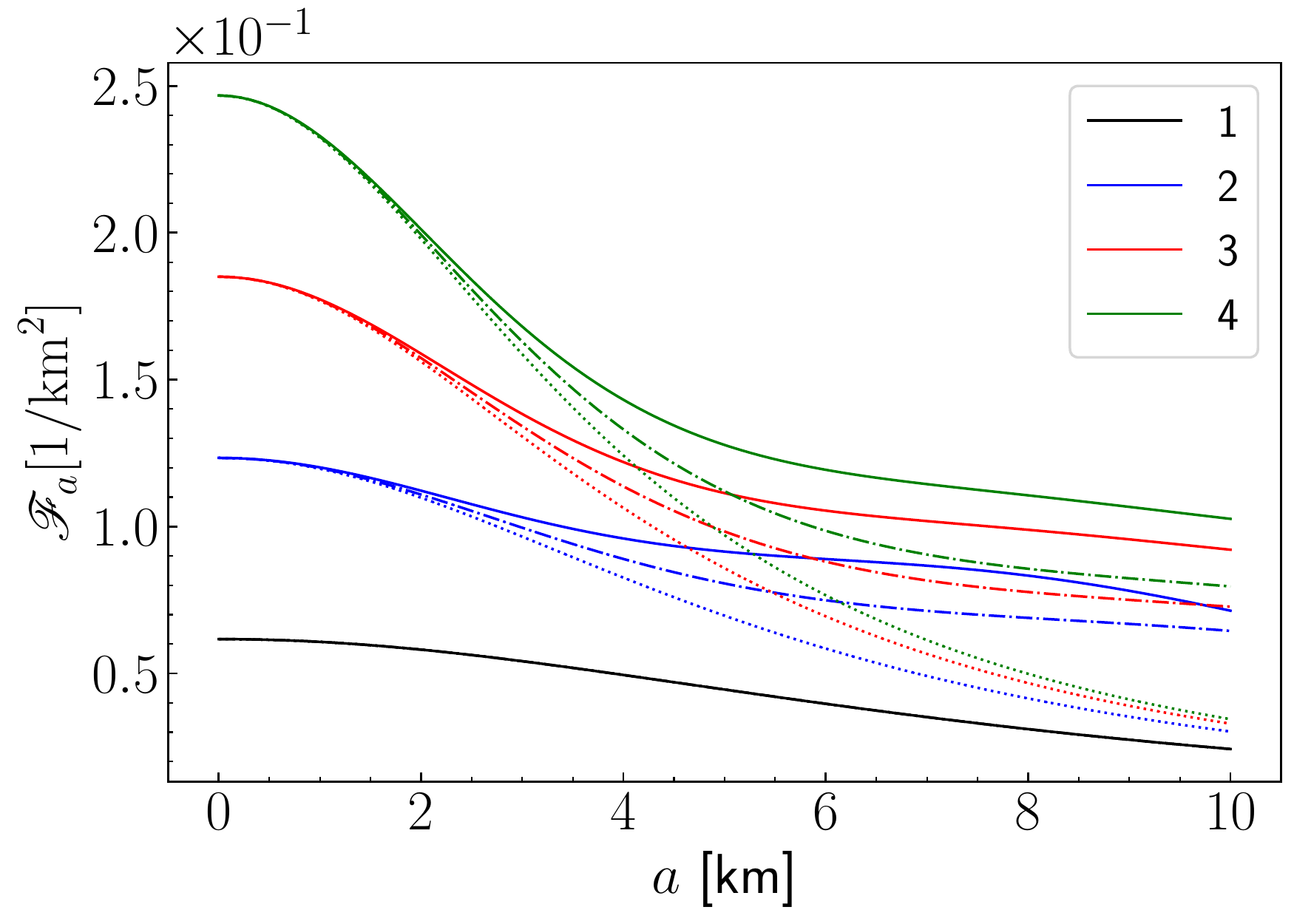}
	\caption{a) The QFI $\mathscr{F}_a$ as a function of $a $ for $(1,2,3,4)$ mode interferometers, which are given by black, blue, red, and yellow lines, respectively. The maximum baseline difference is given by; $\Delta r_{\mathrm{max}} \rightarrow 0$ for dotted lines, $\Delta r_{\mathrm{max}} = 4$ [m] for dashed lines, and $\Delta r_{\mathrm{max}} =6$ [m] for solid lines. The black solid line corresponds to single receiver. $T= 300$ [K].}
	\label{n:a4}
\end{figure}

\textit{Spatial Resolution of Single Circular Source for n mode interferometer array:} To analyze the effect of $n$ for source size estimation, we consider a single circular source as given in Eq.~(\ref{Eqtemp}). In Fig.~(\ref{n:a4}), we show how the QFI for estimating $a$ changes with $n$. For $a \rightarrow 0$, $\mathscr{F}_a $ increases linearly with $n$.
We have $\mathscr{F}_a\sim 6.16 \times 10^{-2} \;\mathrm{[1/km^2]}$ for single receiver which corresponds to a standard deviation of 4 [km] and for higher $n$, we have approximately $\mathscr{F}_a\propto n $ for small values of $a$. If we have an array of 20 receivers, $\mathscr{F}_a \sim 1.23  \;\mathrm{[1/km^2]}$ which gives a standard deviation of 0.9 [km] for estimating $a$. When we increase the source size $a$, we see that there is extra information coming from the phase differences as given by the solid lines for $\Delta r_{\mathrm{max}} =6$ [m] and dashed lines for $\Delta r_{\mathrm{max}} = 4$ [m]. One can also see that as expected the dotted lines, corresponding to the limit $\Delta r_{\mathrm{max}} \rightarrow 0$, get close to the solid black line, which corresponds to a single receiver, for large values of $a$.

To linearly combine these modes, one can calculate the elements of the matrix $\mathbf{M}_i$ numerically. Each normalized eigenvector of $\mathbf{M}_i$ maps to a set of operators $\bar{ \mathbf{d}}$ by linear combination of the operators in $\bar{ \mathbf{b}}$ with corresponding weights and phases. One can design a setup using these weights and phases in the eigenvectors to achieve the resolutions for a chosen parameter given in this section.

\section{Conclusion}\label{conclude}
In summary, we studied possible quantum advantages in passive microwave remote sensing. Starting from a microscopic current density distribution in the source plane corresponding to a position-dependent brightness temperature $T_{\mathrm{eff}}(x,y)$, we derived the general partially coherent state received by an array of receivers. From the dependence of that partially coherent states on parameters that characterize the sources, such as the radius $a$ and brightness temperature $T$ of a uniform circular source, we obtained the quantum Fisher information and hence the quantum Cram\'er-Rao bound for the smallest possible uncertainty with which these parameters can be estimated based on measurements of the multi-mode state of the receivers.  We showed how the optimal measurements allowing one to estimate a single parameter can be obtained for a general antenna array with receivers placed at arbitrary positions.  In general, the optimal measurements correspond to photon-detection in certain detector modes that can be obtained from the original receiver modes by mode mixing via beam-splitters and phase shifters. For single-mode and two-mode interferometers, we gave explicit
analytical results for the best possible resolution of one or two uniform
circular sources, both in  $a$ and $T$ and demonstrated a clear quantum advantage over the classical strategy corresponding to direct heterodyne measurements
of the receiver modes.  In the limit of small source sizes, we recover
known results for the measurement of the centroid and separation of
two-point sources.
We benchmarked our results with the performance
of the SMOS mission, which achieves about 35\,km resolution with 69
antennas deployed on three four-meter long arms arranged in a Y-shape,
operating at 21\,cm
the wavelength,  and flying at a height of
758\,km above Earth. As an example, we showed that by using the
optimal measurements, a single arm of length 4\,m with 20 antennas and a single measurement would allow a spatial resolution of about 1.5\, km. I.e.~with a smaller satellite,
a more than 20-fold increase of resolution compared to SMOS could be
achieved. By increasing the size of the array to 19\,m, the 20
antennas should give rise to a spatial resolution down to 300\,m. {Substantially better resolutions can be achieved if we allow more measurements.  If we assume that the number of samples is given by the time the satellite flies over the object whose size one wants to estimate divided by the inverse bandwidth, even with a single receiver a spatial resolution down to a few meters and a radiometric resolution of a fraction of a Kelvin become possible in principle.} 
\\
Our results generalize previous approaches to quantum-enhanced imaging
based typically on
weak sources (photon numbers on average smaller than 1 per mode) or
point sources, and pave the way to quantum metrological sensitivity
enhancements in real-world scenarios in passive microwave remote
sensing.  Several challenges remain.  Experimentally, single-photon
detection in the microwave regime is still difficult but starts to
become available
\cite{PhysRevX.10.021038}.  It requires very low temperatures for
operating superconducting qubits
that would have to be maintained on a long time scale on the satellite.  From the theoretical side, an extension to a many-parameter regime
requiring adaption of the optimal measurements will be
crucial for true imaging.  Post-measurement beam synthesis that is
common in interferometric astronomy does not work here, as already the
detection modes depend on the pixel in the image that one wants to
focus on.  Nevertheless, the substantial quantum advantages
demonstrated here theoretically in a relatively simple but real-world
scenario give hope that quantum metrology can help to significantly
improve the resolution of passive Earth observation schemes, with
corresponding positive impact on the data available for feeding
climate models, weather forecasts, and forecasts of floodings.

\acknowledgements{DB and EK are grateful for support by the DFG,
  project number BR 5221\textbackslash3-1, DB thanks Yann Kerr, Bernard Roug\'e, and the
  entire SMOS team in Toulouse for
  valuable insights into that mission already a decade ago.}

\onecolumngrid

\appendix

\section{Brightness Temperature and Current Density Fluctuations}
\label{appendixAnew}
The number of photons that pass through a certain receiver area $A_{D}$ in a certain time $t_{D}$ can be found from $\bar{n}=A_{D} t_{D} \Phi$, where $\Phi$ is the photon flux. For a given intensity $I$, the photon flux for frequency $\omega_{0}$ can be found by $\Phi={I}/({\hbar \omega_{0}})$. If the total energy density on the receiver is $U_{D}$, then the intensity can be written as $I=U_{D}c$. Then $\bar n$ becomes $\bar{n}={A_{D} t_{D} U_{D} c}/({\hbar \omega_{0}})$. In the microwave regime $\hbar\omega \ll k_B T $, the energy density (energy per unit volume per frequency) from black body radiation at frequency $\omega$ with a temperature distribution $T(x,y)$ on the surface of radiation at the $i$-th receiver position is given by \cite{braun_fouriercorrelation_2018}
\begin{equation}
	u_{D}(\omega)=\frac{k_{B}}{2 \pi^{3} c^{3}} \int d x d y \omega^{2} \frac{T_{B}(x, y) \cos \tilde \theta(x, y)}{|\mathbf{r}-\mathbf{r}_i|^2},
\end{equation}
where the brightness temperature is defined as $T_{B}(x, y) \equiv T(x, y) B(x, y ; \omega, \tilde \theta, \tilde\varphi)$, Earth is rather a grey than a black body, therefore the emissivity $B(x, y ; \omega, \tilde \theta, \tilde \varphi)$ of the patch in the direction of the satellite given by polar and azimuthal angles is introduced. One can take the integral over $\omega$ using the filter function in Eq.~(\ref{eq:filter}) to find the total energy density (energy per volume) and it becomes 
\begin{equation}
	U_{D}=\frac{k_{B}\omega_0^2 B}{2 \pi^{3} c^{3}} \int d x d y \frac{T_{B}(x, y) \cos \tilde \theta(x, y)}{|\mathbf{r}-\mathbf{r}_i|^2}.
\end{equation}
Then the photon number on the receiver becomes 
\begin{equation}
	\bar n =\frac{2k_{B}}{\pi \hbar \omega_0 } \left(\frac{A_D}{\lambda^2}\right)(t_D B )\int d x d y \frac{T_{B}(x, y) \cos \tilde \theta(x, y)}{|\mathbf{r}-\mathbf{r}_i|^2}.
\label{eqA3}	
\end{equation}
For simplicity of the receivers scattering function, we set $A_D \sim \lambda ^2 $ and $t_D \sim 1/B$. Comparing Eq.~(\ref{eqA3}) with Eq.~(\ref{eq27}), we define $\braket{|\tilde{{j}}_{t,i}\left(\mathbf{r},{\omega}\right)|^2} \equiv K_1 T_{\mathrm{B}}(x,y)\cos\tilde\theta(x,y)\delta(z-R)$ with a constant $K_1 = 32 \tau_c k_B /(3 l_c^3 \mu _0 c)$ which agrees with the result in Ref.~\cite{braun_fouriercorrelation_2018}. 

\section{The general QFI  and the elements of the matrix \textbf{M} for a two-mode interferometer}\label{appendixA}

In this section, we give the general results for the elements of the QFI and the matrix $\mathbf{M}_i$ for a two-mode interferometer, assuming that all the elements of the covariance matrix depend on the parameter $\mu_i$ that we want to estimate. Using the covariance matrix for a two-mode interferometer one finds the QFI matrix elements as
\begin{equation}
    \begin{split}
		\mathscr{F}_{i j} =  &\frac{8}{\mathscr{D}}\left[\partial_{i}\xi^* \partial_{j}\xi ((1-4 \chi^2)^2-4 (1+4 \chi^2) |\xi |^2)\right. +\partial_{i}\xi \partial_{j}\xi^* ((1-4 \chi^2)^2-4 (1+4 \chi^2) |\xi|^2)\\+&4\xi  \partial_{i}\xi^* (\xi\partial_{j}\xi^*  (1+4 \chi^2-4 |\xi|^2)+2\chi \partial_{j}\chi (1-4 \chi^2+4 |\xi |^2))+4 \xi^*\partial_{i}\xi  (\xi\partial_{j}\xi^* (1+4 \chi^2-4 |\xi|^2)\\+&2\chi \partial_{j}\chi (1-4 \chi^2+4 |\xi|^2)) +2 \partial_{i}\chi (-1+4 \chi^2-4 |\xi|^2) (-4 \chi (\xi\partial_{j}\xi^* +\xi^*\partial_{j}\xi )+\left.\partial_{j}\chi (-1+4 \chi^2+4 |\xi |^2))\right],
    \end{split}
\end{equation}
where the denominator is given by
\begin{equation}
    \begin{split}
        \mathscr{D} = (-1+4\chi^2-4|\xi|^2)(16\chi^4+(1-4|\xi|^2)^2-8\chi^2(1+|\xi|^2)).
    \end{split}
    \label{eq:den}
\end{equation}
Using the SLD given in Eq.~(\ref{eq:Sld}) we find the diagonal elements of the matrix $\mathbf{M}_i$ as
\begin{equation}
    \begin{split}
        g^1_i = \frac{2 \left(4 {\partial_i \chi } |\xi|^2+4 {\partial_i \chi } \chi ^2-4 {\partial_i \xi } \xi^* \chi -4 {\partial_i \xi^*} \xi  \chi -{\partial_i \chi }\right)}{16 \chi ^4-8 \chi ^2 (4 |\xi|^2  +1)+(1-4 |\xi|^2)^2},
    \end{split}
\end{equation}
where the two diagonal elements are the same due to the symmetry with respect to the center of the two receivers, and
\begin{equation}
    \begin{split}
        g^2_i &= \frac{2}{\mathscr{D}} \left[-{\partial_i \xi } \left(16 |\xi|^2 \chi ^2+4 |\xi|^2 -16 \chi ^4+8 \chi ^2-1\right)\right.- {\partial_i \xi^*} \left(4 \xi ^2 (4  |\xi|^2-1)-16 \xi ^2 \chi ^2\right)\left.- {\partial_i \chi } \left(32 \xi  \chi ^3-8 \xi  \chi  (4 |\xi|^2+1)\right)\right],
    \end{split}
\end{equation}
where $\mathscr{D}$ is given in Eq.~(\ref{eq:den})

\section{The POVM for heterodyne detection}\label{appendixB}
The POVM for heterodyne detection is given in
Ref.~\cite{tsang_quantum_2011}, and the CFI analyzed for the weak thermal sources. Here we briefly introduce the POVM for heterodyne detection. Then, we compare our results for the QFI with the CFI for heterodyne detection. The POVM is given as
\begin{equation}
    \begin{split}
        E(\nu_1, \nu_2)=\frac{1}{\pi^{2}}|\nu_1, \nu_2\rangle\langle\nu_1, \nu_2|,
    \end{split}
\end{equation}
where $|\nu_1, \nu_2\rangle$ is a coherent state with normalization given by $\int d^{2} \nu_1 d^{2} \nu_2 E(\nu_1, \nu_2)=\mathds{1}$. The covariance matrix for a two-mode interferometer is given in Eq.~(\ref{cov2}). Using the corresponding state for the two-mode interferometer we can find the observation probability for any parameter $\mu_i$, in terms of the elements of the covariance matrix as
\begin{equation}
    \begin{split}
        &P(\nu_1,\nu_2|\mu_i) = \frac{1}{\pi^2((1+\bar{n})^2-|\xi|^2)}\exp\left[{\frac{(-|\nu_1|^2-|\nu_1|^2)(1+\bar{n})+\xi\nu_1^*\nu_2+\xi^*\nu_2^*\nu_1}{(1+\bar{n})^2-|\xi|^2}}\right].
    \end{split}
	\label{eqP}
\end{equation}
The Fisher information for the parameter $\mu_i$ can be found as
\begin{equation}
	\begin{split}
		F_{i}&= \int d^2\nu_1d^2\nu_2 \frac{1}{P(\nu_1,\nu_2|\mu_i)}\left(\frac{\partial P(\nu_1,\nu_2|\mu_i)}{\partial \mu_i}\right)^2\\& =\int d^2\nu_1d^2\nu_2{P(\nu_1,\nu_2|\mu_i)}f(\nu_1,\nu_2)\\&= \braket{f(\nu_1,\nu_2)},
	\end{split}
	\label{eqF}
\end{equation}
where $f(\nu_1,\nu_2)$ is a polynomial function of second and fourth order correlations of $\nu_1$ and $\nu_2$, defined as
\begin{equation}
	\begin{split}
		f(\nu_1,\nu_2) \equiv \left(\partial_{\mu_i} \log\left(P(\nu_1,\nu_2|\mu_i) \right)\right)^2 =\frac{1}{(P(\nu_1,\nu_2|\mu_i))^2}\left(\frac{\partial P(\nu_1,\nu_2|\mu_i)}{\partial \mu_i}\right)^2.
	\end{split}
	\label{eqf}
\end{equation}
With Wick's theorem for Gaussian states, the fourth order statistic can be written as
\begin{equation}
	\begin{split}
		\braket{x_1 x _2x _3x_4} = \braket{x_1x_2}\braket{x_3x_4} +   \braket{x_1x_3}\braket{x_2x_4} + \braket{x_1x_4}\braket{x_2x_3} ,
	\end{split}
\end{equation}
where $x_i \in \{\nu_1,\nu_1^*,\nu_2,\nu_2^*\} $. We can also write $\braket{|{\nu_1}|^2} =\braket{|{\nu_2}|^2} = 1+\bar{n} $ and $\braket{\nu_1^*\nu_2} =\xi $, $\braket{\nu_2^*\nu_1} =\xi^* $.

\section{The uniform circular source for a two-mode interferometer} \label{appendixC}
We find the elements of the covariance matrix describing the state of two-mode interferometers in Eq.~(\ref{cov2}). Then for a circular source with size $a$ located at position $(x_0,y_0,R)$ with the assumption $x_0,y_0 \ll R$ in the source plane we have
\begin{equation}
\begin{split}
\bar{n} &=\frac{\kappa T}{R^2}\int dxdy \mathrm{circ}({x-x_0,y-y_0}), \\&= \frac{\pi a^2 \kappa T}{R^2},
\end{split}
\label{ncirc}
\end{equation}
and $\chi$ and $\xi$  become
\begin{equation}
\begin{split}
\chi = \frac{1}{2}+ \frac{\pi a^2 \kappa T}{R^2},
\end{split}
\label{xicirc}
\end{equation}

\begin{equation}
\begin{split}
        \xi = \braket{b_2^\dagger b_1} &= \frac{\kappa T}{R^2}\int dxdy \mathrm{circ}({x-x_0,y-y_0})\exp\left( 2\pi i( x v_x + y v_y )\right) \\& = \frac{\kappa T a^2}{R^2} \frac{J_1\left(2\pi a \sqrt{v_x^2 +v_y^2} \right)}{a \sqrt{v_x^2 +v_y^2}} \exp\left( 2\pi i( x_0 v_x + y_0 v_y )\right) ,
\end{split}
\end{equation}
where $v_{x} = \Delta r \cos\varphi /(\lambda R)$, $v_{y} = \Delta r \sin\varphi /(\lambda R)$, with $\mathbf{\Delta r} = \Delta r(\cos \varphi, \sin \varphi,0) $. Note that $\sqrt{v_x^2 +v_y^2} = \Delta r /(\lambda R)$.

\textit{\\The Quantum Fisher Information: The uniform circular source \\}

We found the QFI for estimating $a$ is as
\begin{equation}
    \begin{split}
        \mathscr{F}_a =& \frac{8 \pi ^2 a \Delta r^2 \kappa  T }{\mathscr{D}_a}\left[\pi  a {\Delta r}^2 (\pi  a^2 \kappa  T+R^2) (J_0^2+1)-2 {\Delta r} \lambda  R (2 \pi  a^2 \kappa  T+R^2) J_0 J_1+ a \kappa  \lambda ^2 R^2 T (J_0^2+1) J_1^2\right],
    \end{split}
	\label{QFIa}
\end{equation}
where
\begin{equation}
    \begin{split}
       \mathscr{D}_a&= \left(\pi ^2 a^2 \Delta r^2-\lambda ^2 R^2 J_1^2\right)\left(\Delta r^2 \left(\pi  a^2 \kappa  T+R^2\right)^2-\kappa ^2 \lambda ^2 R^2 T^2 J_1^2\right),
    \end{split}
\end{equation}
and $J_i \left(\frac{2 a \Delta r  \pi }{R \lambda }\right)$ are the Bessel functions of the first kind and $i$-th order. The QFI for estimating $T$ becomes
\begin{equation}
    \begin{split}
        \mathscr{F}_{T} &=\frac{2 \kappa  a^2 \left(\pi  \Delta r ^2 \left(\pi  a^2 \kappa  T+R^2\right)-\kappa  \lambda ^2 R^2 T J_1^2\right)}{T \left(\Delta r ^2 \left(\pi  a^2 \kappa  T+R^2\right)^2-a^2\kappa ^2 \lambda ^2 R^2 T^2 J_1{}^2\right)}\,.
    \end{split}
\end{equation}
The other elements regarding the source size and the temperature of the circular source can be found as
\begin{equation}
	\begin{split}
		\mathscr{F}_{aT} = \frac{4 \pi  a {\Delta r} \kappa  \left({\Delta r} \left(\pi  a^2 \kappa  T+R^2\right)-a \kappa  \lambda  R T J_0 J_1\right)}{{\Delta r}^2 \left(\pi  a^2 \kappa  T+R^2\right)^2-a^2 \kappa ^2 \lambda ^2 R^2 T^2 J_1{}^2}.
	\end{split}
	\label{QFIaT}
\end{equation}
The QFI matrix elements for estimating the source locations can be written as
\begin{equation}
	\begin{split}
		\mathscr{F}_{i_0j_0} = \frac{8 \pi ^2 R^2 \lambda^2  \kappa  T J_1{}^2 v_i v_j}{\pi  {\Delta r}^2 \left(\pi  a^2 \kappa  T+R^2\right)-\kappa  \lambda ^2 R^2 T J_1{}^2},
	\end{split}
\end{equation}
where $i,j \in \{x,y\}$.

\textit{\\The elements of the matrix $\mathbf{M}_i$ for a two-mode interferometer: The uniform circular source \\}

To combine two modes of the receivers for the optimum measurements, we calculate $\delta$ as given in Eq.\eqref{eq:V}. We find the matrix elements of $\mathbf{M}_a$ as
\begin{equation}
	\begin{split}
		g^1_a &= \frac{2\pi  {\Delta r}^2 R^2 }{\mathscr{D}_a}(\pi  a {\Delta r}^2 (\pi  a^2 \kappa  T+R^2)+\lambda  R J_1 (a \kappa  \lambda  R T J_1-{\Delta r} (2 \pi  a^2 \kappa  T+R^2) J_0)),
	\end{split}
	\label{ga1}
\end{equation}
\begin{equation}
	\begin{split}
		g^2_a &= \frac{2\pi  {\Delta r}^2 R^2 }{\mathscr{D}_a}(a J_0 (\pi  {\Delta r}^2 (\pi  a^2 \kappa  T+R^2)+\kappa  \lambda ^2 R^2 T J_1{}^2) -{\Delta r} \lambda  R (2 \pi  a^2 \kappa  T+R^2) J_1)e^{-i\delta},
	\end{split}
	\label{ga2}
\end{equation}
where $\delta = v_x x_0 + v_y y_0$. For the temperature estimation we get the elements of $\mathbf{M}_T$ as 
\begin{equation}
	\begin{split}
		g^1_T &=\frac{ {\Delta r}^2 R^2 \left(\pi  a^2 \kappa  T+R^2\right)}{ {\Delta r}^2 T \left(\pi  a^2 \kappa  T+R^2\right)^2- a^2 \kappa ^2 \lambda ^2 R^2 T^3 J_1^2},
	\end{split}
\end{equation}
\begin{equation}
	\begin{split}
		g^2_T &=-\frac{a  {\Delta r} \kappa  \lambda  R^3 J_1e^{-i \delta } }{a^2 \kappa ^2 \lambda ^2 R^2 T^2 J_1^2- {\Delta r}^2 \left(\pi  a^2 \kappa  T+R^2\right)^2}.
	\end{split}
\end{equation}
Finally, for the source location we found
\begin{equation}
	\begin{pmatrix}
		g^2_{x_0}  \\
		g^2_{y_0}
		\end{pmatrix} =-\frac{ 2\pi  {\Delta r}^2 R^2 J_1 e^{i(-\delta +\pi/2 )}}{a \left(\pi  {\Delta r}^2 \left(\pi  a^2 \kappa  T+R^2\right)-\kappa  \lambda ^2 R^2 T J_1^2\right)}\begin{pmatrix}
			\cos (\varphi )   \\
			\sin (\varphi )
		\end{pmatrix},
	\label{g2ij}
\end{equation}
and $g^1_{x_0}=g^1_{y_0}= 0$.

\textit{\\The classical Fisher information for heterodyne detection: The Uniform Circular Source \\}

Since we calculated the elements of the covariance matrix in Eq.~(\ref{ncirc}) and Eq.~(\ref{xicirc}) we can calculate Eq.~(\ref{eqP}) and Eq.~(\ref{eqf}). Using the CFI for the heterodyne detection in Eq.~(\ref{eqF}), we can write the result for estimating the source size as
\begin{equation}
	\begin{split}
		F_a=&\frac{8 \pi ^2 a^2 \kappa ^2 T^2 \Delta r^3(\pi  a^2 \kappa  T+R^2) }{D_a}[4 a^5 \kappa ^5 \lambda ^5 R^5 T^5 J_0 J_1{}^5-2 a^2 {\Delta r}^3 \kappa ^2 \lambda ^2 R^2 T^2 (\pi  a^2 \kappa  T+R^2)^3 (J_0{}^2+1) J_1{}^2\\&+{\Delta r}^5 (\pi  a^2 \kappa  T+R^2)^5 (J_0{}^2+1)-4 a {\Delta r}^4 \kappa  \lambda  R T (\pi  a^2 \kappa  T+R^2)^4 J_0 J_1\\&-7 a^4 {\Delta r} \kappa ^4 \lambda ^4 R^4 T^4 (\pi  a^2 \kappa  T+R^2) (J_0{}^2+1) J_1{}^4+16 a^3 {\Delta r}^2 \kappa ^3 \lambda ^3 R^3 T^3 (\pi  a^2 \kappa  T+R^2)^2 J_0 J_1{}^3],
	\end{split}
	\label{CFIa}
\end{equation}
with
\begin{equation}
	\begin{split}
		D_a={ (\Delta r^2(\pi  a^2 \kappa  T+R^2)^2-{a^2 \kappa ^2 \lambda ^2 R^2 T^2 J_1{}^2}){}^4}.
	\end{split}
\end{equation}
The estimation of the temperature becomes
\begin{equation}
	\begin{split}
		F_T &= \frac{2 a^2 {\Delta r}^2 \kappa ^2 (\pi  a^2 \kappa  T+R^2)}{D_T} (\pi ^2 a^2 {\Delta r}^6 (\pi  a^2 \kappa  T+R^2)^5-a^4 \kappa ^4 \lambda ^6 R^6 T^4 (3 \pi  a^2 \kappa  T+7 R^2) J_1{}^6\\&+{\Delta r}^4 \lambda ^2 R^2 (\pi  a^2 \kappa  T+R^2)^3 (-5 \pi ^2 a^4 \kappa ^2 T^2-2 \pi  a^2 \kappa  R^2 T+R^4) J_1{}^2+a^2 {\Delta r}^2 \kappa ^2 \lambda ^4 R^4 T^2 (7 \pi ^3 a^6 \kappa ^3 T^3\\&+19 \pi ^2 a^4 \kappa ^2 R^2 T^2+10 \pi  a^2 \kappa  R^4 T-2 R^6) J_1{}^4),
	\end{split}
\end{equation}
where
\begin{equation}
	\begin{split}
		D_T= {\left({\Delta r}^2 \left(\pi  a^2 \kappa  T+R^2\right)^2-a^2 \kappa ^2 \lambda ^2 R^2 T^2 J_1{}^2\right){}^4}.
	\end{split}
\end{equation}

\section{Two point sources for a two-mode interferometer}\label{appendixD}
\begin{figure}[h!]
	\centering
	\includegraphics[width=0.5\linewidth]{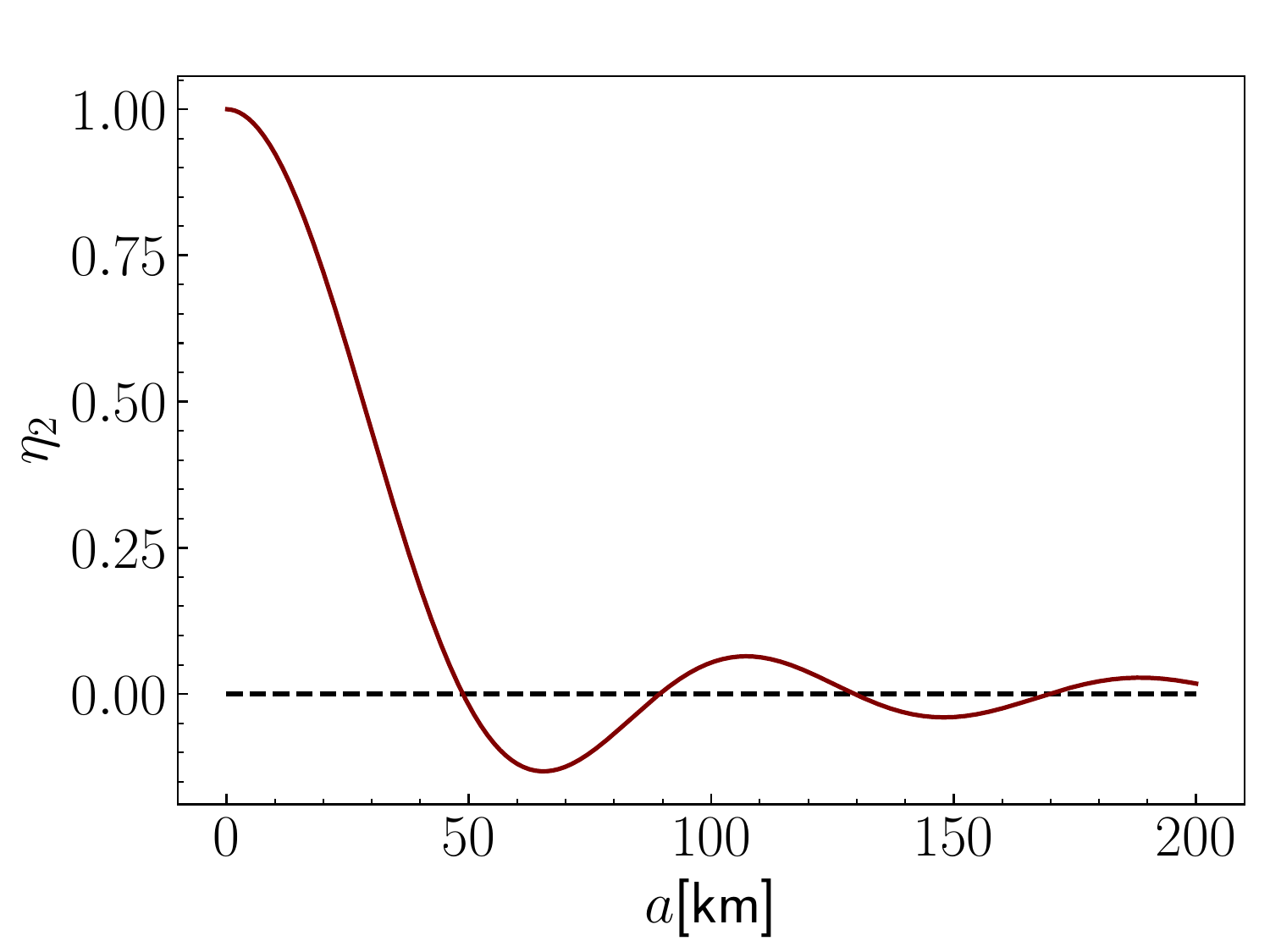}
	\caption{Plot showing the behavior of $\eta_2$ with respect to the radius of the circular disc source.}
	\label{bes}
\end{figure}

The temperature distribution of two circular sources with equal size $a$ at locations $(x_1,y_1,R)$ and $(x_2,y_2,R)$ is given in Eq.~(\ref{twotemp}). We assume that $\{|x_i|,|y_i|,a\} \ll R $. The elements of the covariance matrix in Eq.~(\ref{cov2}) for two point sources with different temperature can be found using Eq.~(\ref{eqn}), Eq.~(\ref{eqxi}) and Eq.~(\ref{twotemp}) as
\begin{equation}
\begin{split}
\chi &= \frac{1}{2}+ \sum_i\frac{\pi a^2 \kappa T_i}{R^2} = \frac{1}{2}+ \frac{2\pi a^2 \kappa T}{R^2} \\& = \frac{1}{2}+ 2\eta \kappa T ,
\end{split}
\end{equation}
where $\eta = \pi a^2 /R^2 $, and
\begin{equation}
\begin{split}
\xi = \braket{b^\dagger_1 b_2} &= \frac{\kappa \pi a^2  }{R^2} \frac{2{J_1\left(\frac{2\pi a \Delta r }{R\lambda}\right)} }{\frac{2\pi a \Delta r }{R\lambda}} (T_1 e^{2\pi i(v_xx_1+v_yy_1) }+T_2 e^{2\pi i(v_xx_2+v_yy_2)  }) \\& = \frac{\kappa\eta \eta_2}{2} ((2T-\Delta T) e^{2\pi i(v_xx_1+v_yy_1)}+(2T+\Delta T)  e^{2\pi i(v_xx_2+v_yy_2)}),
\end{split}
\end{equation}
where the average temperature is defined as $T\equiv(T_1+T_2)/2$, and the temperature difference of the sources as $\Delta T \equiv T_2 - T_1 $ with $T_2 \geqslant T_1$ assumed, while the parameter $\eta_2$ is given by
\begin{equation}
    \eta_2 = \frac{2{J_1\left(\frac{2\pi a \Delta r }{R\lambda}\right)} }{\frac{2\pi a \Delta r }{R\lambda}},
\end{equation}
which is related to the source size. In Fig.~(\ref{bes}), we can see the behavior of $\eta_2$ with respect to the source size. For point sources one can approximate $\eta_2 \approx 1 $. \\\\

\textit{\\The Quantum Fisher Information: Two Point Sources\\}

We found the elements of the QFI matrix for estimating the source separation as
 \begin{equation}
	 \begin{split}
		\mathscr{F}_{s_i s_j}&=\frac{2 \pi ^2 \eta  \kappa  {v_i} {v_j}}{\mathscr{D}_{s_xs_y}} (\eta ^2 \kappa ^2 (4 T^2-{\Delta T}^2) ((4 T^2-{\Delta T}^2) \cos (4 \pi  ({s_x} {v_x}+{s_y} {v_y}))+16 T^2 \cos (2 \pi  ({s_x} {v_x}+{s_y} {v_y})))\\&-\eta ^2 \kappa ^2 ({\Delta T}^4-24 {\Delta T}^2 T^2+80 T^4)-128 \eta  \kappa  T^3-32 T^2),
	 \end{split}
 \end{equation}
where $i,j= \{x,y\}$ and the denominator is given by
\begin{equation}
	\begin{split}
		\mathscr{D}_{s_xs_y}&=\eta  \kappa  (4 T^2-{\Delta T}^2) (\eta ^2 \kappa ^2 ({\Delta T}^2-4 T^2) \cos (4 \pi  ({s_x} {v_x}+{s_y} {v_y}))\\&+4 (\eta  \kappa  (-{\Delta T}^2 \eta  \kappa +4 \eta  \kappa  T^2+6 T)+1) \cos (2 \pi  ({s_x} {v_x}+{s_y} {v_y})))\\&-3 \eta ^3 \kappa ^3 ({\Delta T}^2-4 T^2)^2+24 \eta ^2 \kappa ^2 T ({\Delta T}^2-4 T^2)+4 \eta  \kappa  ({\Delta T}^2-20 T^2)-16 T\,.
	\end{split}
\end{equation}
The elements of QFI matrix for estimating the centroid becomes
 \begin{equation}
	 \begin{split}
		 \mathscr{F}_{t_i t_j} &= \frac{16\pi^2 v_i v_j \eta  \kappa  }{\mathscr{D}_t} [(4 T^2-{\Delta T}^2) \cos \left(2\pi ({s_x} v_x+s_yv_y)\right)+{\Delta T}^2+4 T^2],
	 \end{split}
	 \label{QFIt}
 \end{equation}
 where the denominator is
 \begin{equation}
	 \begin{split}
		 \mathscr{D}_t &= 4 T+4 \eta  \kappa  T^2-{\Delta T}^2 \eta  \kappa -\eta  \kappa  (4 T^2-{\Delta T}^2) \cos \left(2\pi ({s_x} v_x+s_yv_y)\right).
	 \end{split}
 \end{equation}
Off diagonal elements of the QFI matrix can be found as
\begin{equation}
	\begin{split}
		\mathscr{F}_{s_it_j} = \frac{32 \pi ^2 {\Delta T} \eta  \kappa  T {v_i} {v_j}}{{\Delta T}^2 \eta  \kappa +\eta  \kappa  \left(4 T^2-{\Delta T}^2\right) \cos (2 \pi  ({s_x} {v_x}+{s_y} {v_y}))-4 \eta  \kappa  T^2-4 T}.
	\end{split}
\end{equation}
If we align two receivers parallel to the source separation, $v_x \rightarrow v $ and $v_y \rightarrow 0 $. In the limit where $\Delta T \rightarrow 0$ the QFI for the source separation simplifies to
 \begin{equation}
     \begin{split}
        \mathscr{F}_s \rightarrow
         \frac{4\pi^2  v^2 \eta  \kappa  T \left(\eta  \kappa  T \cos \left(2 \pi  {s} v\right)+3 \eta  \kappa  T+1\right)}{\left(1+4 \eta  \kappa  T+2 \eta ^2 \kappa ^2 T^2-2 \eta ^2 \kappa ^2 T^2 \cos \left(2 \pi  {s} v\right)\right)},
     \end{split}
 \end{equation}
 and the QFI for the centroid simplifies to
 \begin{equation}
    \begin{split}
        \mathscr{F}_t \rightarrow \frac{32\pi^2 v^2\eta  \kappa  T  \cos ^2\left(\pi sv\right)}{1+\eta  \kappa  T -\eta  \kappa  T \cos \left(2\pi sv\right)},
    \end{split}
\end{equation}
 which agrees with the results in Ref.~\cite{wang_superresolution_2021} for ($\Delta T = 0,v_y = 0, s_y = 0 ,t_y = 0$) as expected.

 \textit{\\The elements of the matrix $\mathbf{M}_i$ for a two-mode interferometer: Two Point Sources \\} 
 
 For simplicity let us assume that $\Delta T\rightarrow 0 $. Then we have the elements of the matrix $\mathbf{M}_i$ for a two-mode interferometer for estimating the source sizes as
\begin{equation}
	\begin{split}
		g^1_{s_i} &=\frac{\pi  v_i (4 \eta  \kappa  T+1)\cot (\pi  (s_xv_x +s_yv_y))}{ (1+4 \eta  \kappa  T+2 \eta ^2 \kappa ^2 T^2-2 \eta ^2 \kappa ^2 T^2 \cos (2 \pi  (s_xv_x +s_yv_y)))}, \\
		g^2_{s_i} &=\frac{\pi  v_i  (\eta  \kappa  T \cos (2 \pi  (s_xv_x +s_yv_y))+3 \eta  \kappa  T+1)\csc (\pi  (s_xv_x +s_yv_y))\exp({-i \delta_s}) }{ (1+4 \eta  \kappa  T+2 \eta ^2 \kappa ^2 T^2-2 \eta ^2 \kappa ^2 T^2 \cos (2 \pi  (s_xv_x +s_yv_y)))},
	\end{split}
\end{equation}
and
\begin{equation}
	\begin{split}
        g^1_{t_i} &= 0, \\
		g^2_{t_i} &=\frac{2\pi  v  \cos (\pi  (s_xv_x +s_yv_y))\exp({-i\delta_t})}{1+\eta  \kappa  T-\eta  \kappa  T \cos (2 \pi  (s_xv_x +s_yv_y))},
	\end{split}
\end{equation}
where $\delta_s = 2\pi (t_x v_x+t_yv_y)-\pi$
 and $\delta_t = 2\pi (t_x v_x+t_yv_y)+\pi/2$.

\textit{\\The classical Fisher information for heterodyne detection: Two point sources \\}

Using the CFI given for the heterodyne detection in Eq.~(\ref{eqF}), and assuming that both sources have the same temperature ($\Delta T\rightarrow 0$), one can find the CFI for estimating the centroid as
\begin{equation}
	\begin{split}
		F_{t_i} &=\frac{32 \pi ^2 \eta ^2 \kappa ^2 T^2 {v_i}^2 (2 \eta  \kappa  T+1)^2 \cos ^2(\pi  ({s_x} {v_x}+{s_y} {v_y}))}{\left(-2 \eta ^2 \kappa ^2 T^2 \cos (2 \pi  ({s_x} {v_x}+{s_y} {v_y}))+2 \eta  \kappa  T (\eta  \kappa  T+2)+1\right)^2},
	\end{split}
	\label{CFIt}
\end{equation}
 Again assuming ($\Delta T\rightarrow 0 $), we can find the CFI for estimating the source separation as
\begin{equation}
	\begin{split}
		F_{s_i} &= \frac{1}{D_s}{8 \pi ^2 \eta ^2 \kappa ^2 T^2 v_i^2 \sin ^2(\pi  (s_xv_x +s_yv_y)) (2 \eta  \kappa  T+1)^2}  [1- 14 \eta ^4 \kappa ^4 T^4 \cos (4 \pi  (s_xv_x +s_yv_y))\\&-4 \eta ^2 \kappa ^2 T^2 \cos (2 \pi  (s_xv_x +s_yv_y)) (2 \eta  \kappa  T (9 \eta  \kappa  T+2)+1)-2 \eta  \kappa  T (\eta  \kappa  T (\eta  \kappa  T (21 \eta  \kappa  T-8)-10)-4)],
	\end{split}
	\label{CFIs}
\end{equation}
where the denominator is given by
\begin{equation}
	\begin{split}
		D_{s}=(1+4 \eta  \kappa  T+2 \eta ^2 \kappa ^2 T^2-2 \eta ^2 \kappa ^2 T^2 \cos (2 \pi  (s_xv_x +s_yv_y)))^4.
	\end{split}
\end{equation}

\twocolumngrid
\newpage
\vfill
\bibliography{bibliography}

\end{document}